\documentclass[nofootinbib,amsmath,onecolumn,notitlepage,preprintnumbers,floatfix]{revtex4-1}
\usepackage{multirow}
\usepackage{amssymb, esvect, amsmath, graphicx, latexsym, amsthm, slashed, eso-pic}
\usepackage{xcolor}
\usepackage{hyperref}
\usepackage{amsmath}

 \def\d{{\rm d}}   
       
\newcommand{\beq}{\begin{equation}} \newcommand{\eeq}{\end{equation}}
\newcommand{\bea}{\begin{eqnarray}} \newcommand{\eea}{\end{eqnarray}}

\def\lsim{\mathrel{\raise.3ex\hbox{$<$\kern-.75em\lower1ex\hbox{$\sim$}}}}
\def\gsim{\mathrel{\raise.3ex\hbox{$>$\kern-.75em\lower1ex\hbox{$\sim$}}}}

\newcommand{\be}{\begin{eqnarray}}
\newcommand{\ee}{\end{eqnarray}}

\newcommand{\benum}{\begin{enumerate}}
\newcommand{\eenum}{\end{enumerate}}
\newcommand{\bi}{\begin{itemize}}
\newcommand{\ei}{\end{itemize}}

\begin{document}

\preprint{FERMILAB-PUB-23-038-T, LAPTH-005/23}

\title{Signals of a New Gauge Boson from IceCube and Muon $g-2$}

\author{Dan Hooper$^{a,b,c}$}
\thanks{ORCID: http://orcid.org/0000-0001-8837-4127}

\author{Joaquim Iguaz Juan$^{d}$}
\thanks{ORCID: http://orcid.org/0000-0002-7203-5216}

\author{Pasquale D. Serpico$^{d}$}
\thanks{ORCID: http://orcid.org/0000-0002-8656-7942}

\affiliation{$^a$University of Chicago, Kavli Institute for Cosmological Physics, Chicago IL, USA}
\affiliation{$^b$University of Chicago, Department of Astronomy and Astrophysics, Chicago IL, USA}
\affiliation{$^c$Fermi National Accelerator Laboratory, Theoretical Astrophysics Group, Batavia, IL, USA}
\affiliation{$^d$LAPTh,  CNRS, USMB, F-74940 Annecy, France}
\date{\today}

\begin{abstract}
A $Z'$ boson associated with a broken $U(1)_{L_{\mu} - L_{\tau}}$ gauge symmetry offers an economical solution to the long-standing $g_\mu-2$ anomaly, confirmed and strengthened by recent measurements at Fermilab. Here, we revisit the impact of such a $Z'$ on the spectrum of high-energy astrophysical neutrinos, as measured by the IceCube experiment. This spectrum has been observed to exhibit a dip-like feature at $E_{\nu} \sim 0.2-1 \, {\rm PeV}$, which could plausibly arise from the physics of the sources themselves, but could also be the consequence of high-energy neutrinos resonantly scattering with the cosmic neutrino background, mediated by a $Z'$ with a mass on the order of $m_{Z'} \sim 10 \, {\rm MeV}$. In this study, we calculate the impact of such a $Z'$ on the high-energy neutrino spectrum for a variety of model parameters and source distributions. For couplings that can resolve the $g_{\mu}-2$ anomaly, we find that this model could self-consistently produce a spectral feature that is consistent with IceCube's measurement, in particular if the neutrinos observed by IceCube predominantly originate from high-redshift sources.

\end{abstract}

\maketitle

\section{Introduction}

Measurements of the anomalous muon's magnetic moment, $a_{\mu} \equiv (g_{\mu}-2)/2$, performed at Fermilab~\cite{Muong-2:2021ojo,Muong-2:2021vma} and at the Brookhaven National Laboratory~\cite{Muong-2:2006rrc} have yielded an experimental average of $a_{\mu}^{\rm EXP} =116592061(41) \times 10^{-11}$. Comparing this to the value predicted by the Standard Model based on dispersion relations~\cite{Aoyama:2020ynm}, one obtains
\begin{align}
\Delta a_{\mu} \equiv a^{\rm EXP}_{\mu} - a^{\rm SM}_{\mu} = 251 (59) \times 10^{-11},
\end{align}
constituting a $4.2\sigma$ discrepancy. In the years ahead, we expect the experimental uncertainties associated with this measurement to be reduced considerably. In tandem, studies of the hadronic contributions to $a_{\mu}$ using lattice QCD techniques~\cite{Davier:2019can,Blum:2019ugy,Chao:2021tvp,Borsanyi:2020mff,Bazavov:2023has,Blum:2023qou}, which currently hint at a lower significance for this discrepancy, promise to substantially refine the Standard Model prediction for this quantity.

A variety of scenarios involving new physics have been proposed to potentially resolve this discrepancy (for reviews, see Refs.~\cite{Athron:2021iuf,Capdevilla:2021rwo}). From among these possibilities, perhaps the simplest class of models are those which introduce a new particle with an MeV-scale mass that couples to muons with a strength on the order of $g_{\mu} \sim 10^{-4}$~\cite{Pospelov:2008zw,Chen:2017awl,Fayet:2007ua}. Such a particle could have significant implications for astrophysics and cosmology, opening up the possibility that such a state could be constrained or studied using astrophysical probes~\cite{Escudero:2019gzq,Holst:2021lzm}. In particular, if such a particle existed, it could cause high-energy neutrinos to appreciably scatter with the cosmic neutrino background, impacting the propagation of high-energy neutrinos across cosmological distance scales. Such interactions could induce spectral features that would be measurable at large-volume neutrino telescopes such as IceCube~\cite{Hooper:2007jr,Araki:2014ona,Kamada:2015era,DiFranzo:2015qea,Carpio:2021jhu}.

The IceCube Collaboration has reported the detection of an approximately isotropic flux of astrophysical neutrinos, spanning energies between several TeV and several PeV~\cite{IceCube:2020acn,IceCube:2021rpz,IceCube:2013cdw,IceCube:2013low,IceCube:2014stg}. Neutrino events observed at IceCube can be classified as either muon tracks from $\nu_{\mu}$ charged current interactions, or showers from all other flavors and interactions. In this study, we focus on the 6-year dataset of shower events presented in Ref.~\cite{IceCube:2020acn}, as such events allow for the most direct measurement of the underlying neutrino spectrum. We will consider how this spectrum might be altered in models which include a MeV-scale gauge boson with couplings motivated by the $g_{\mu}-2$ anomaly.

This article is structured as follows. In Sec.~\ref{model} we review the model under consideration and its impact on the propagation and spectrum of high-energy neutrinos. In Sec.~\ref{fluximpact}, we study the implications for the flux measured at IceCube, first using simplified models for the source distribution (Sec.~\ref{s2}), then for more realistic astrophysical redshift distributions (Sec.~\ref{s3}).  In Sec.~\ref{dark}, we consider models that include extra states which reside within a dark sector, to which the new gauge boson acts as a portal. Such a scenario could quite plausibly include a candidate for the dark matter of our universe, and could give rise to a rich variety of neutrino phenomenology. Finally, in Sec.~\ref{conclusions}, we summarize our results and discuss directions for future research.

\section{A new MeV-scale gauge boson and high-energy neutrinos}\label{model}

Gauge symmetries beyond those of the Standard Model are a feature of many scenarios involving new physics~\cite{Langacker:2008yv}. In particular, new broken abelian $U(1)$ gauge symmetries, which give rise to the existence of a massive $Z'$ boson, can arise within the context of Grand Unified Theories~\cite{London:1986dk,Hewett:1988xc}, little Higgs theories~\cite{Arkani-Hamed:2002iiv,Han:2003wu,Perelstein:2005ka}, dynamical symmetry breaking scenarios~\cite{Hill:2002ap}, models with extra spatial dimensions~\cite{Agashe:2003zs,Agashe:2007ki,Carena:2003fx}, string inspired models~\cite{Braun:2005bw,Cleaver:1998gc,Lebedev:2007hv,Cvetic:2001nr}, and many other proposed extensions of the Standard Model~\cite{Arkani-Hamed:2001nha,Cvetic:1997ky,Langacker:1999hs}.

The phenomenology of a $Z'$ boson depends on its mass, the strength of its gauge coupling, and on which particles are charged under its corresponding gauge symmetry. While there are many examples of $U(1)$ symmetries that could be manifest in nature, most of these possibilities require the introduction of new chiral fermions (known as ``exotics'') to cancel gauge anomalies~\cite{Batra:2005rh,Appelquist:2002mw}. From the criteria of simplicity, a $U(1)$ which does not require any such exotics would be particularly attractive. As it turns out, the only anomaly-free $U(1)$ models are those which gauge baryon-minus-lepton number ($B-L$), the difference of two lepton flavors ($L_i-L_j$), baryon number minus three units of one lepton flavor ($B-3L_i$), or a quantity that does not involve any Standard Model charges.

In the light of the very stringent constraints that have been placed on the couplings of a light $Z'$ to electrons or light quarks, the only anomaly-free $U(1)$ that could potentially explain the observed $g_{\mu}-2$ anomaly is one that gauges the quantity $L_{\mu} - L_{\tau}$~\cite{He:1990pn,He:1991qd}. After the spontaneous breaking of this symmetry, the Lagrangian for this model is given by
\begin{align}
\label{eq:lag}
\mathcal{L} = \mathcal{L}_{\rm SM} - \frac{1}{4} Z'^{\alpha \beta} Z'_{\alpha \beta} + \frac{m_{Z'}^2}{2} Z'_{\alpha} Z'^{\alpha} + Z'_{\alpha} J^{\alpha}_{\mu-\tau},
\end{align}
where $\mathcal{L}_{\rm SM}$ is the Standard Model Lagrangian, $Z'^{\alpha \beta} \equiv \partial_{\alpha} Z'_{\beta} - \partial_{\beta} Z'_{\alpha}$ is the field strength tensor, and $m_{Z'}$ is the mass of the new gauge boson. If no new states charged under this symmetry exist, the $\mu-\tau$ current is given by
\begin{align}
J^{\alpha}_{\mu-\tau} = g_{Z'} \, (\bar{\mu} \gamma^{\alpha} \mu + \bar{\nu}_{\mu} \gamma^{\alpha} P_L \nu_{\mu} - \bar{\tau} \gamma^{\alpha} \tau - \bar{\nu}_{\tau} \gamma^{\alpha} P_L \nu_{\tau}),
\end{align}
where $g_{Z'}$ is the new gauge coupling and $P_L \equiv (1-\gamma_5)/2$.

The $Z'$ associated with a $U(1)_{L_{\mu} - L_{\tau}}$ gauge group will lead to the following correction to the muon's magnetic moment (at leading order in terms of powers of $g_{Z'}$)~\cite{Jegerlehner:2009ry}:
\begin{align}
\Delta a_{\mu} = \frac{g^2_{Z'} m^2_{\mu}}{4 \pi^2 m^2_{Z'}} \int^1_0 \frac{x^2 (1-x) dx}{1-x+(m^2_{\mu}/m^2_{Z'})x^2}.
\end{align}
Such a contribution could accommodate the measured value of the muon's magnetic moment for $m_{Z'} \sim 10-300 \, {\rm MeV}$; below this range, such $Z'$s are ruled out by cosmological considerations~\cite{Escudero:2019gzq}, while larger masses are excluded by laboratory constraints~\cite{CCFR:1991lpl,Altmannshofer:2014pba,BaBar:2016sci}.

The existence of a new MeV-scale gauge boson with couplings motivated by the $g_{\mu}-2$ anomaly would lead to a significant cross section for neutrino-neutrino scattering. In the presence of such an interaction, the scattering of high-energy neutrinos with the cosmic neutrino background could induce potentially observable features in the astrophysical spectrum of such particles. 

The final term in Eq.~\eqref{eq:lag} leads to the following cross section for neutrinos of mass eigenstates, $i$ and $j$:
\begin{align}
\label{sigma}
\sigma(\nu_i \bar{\nu}_j \rightarrow \nu \bar{\nu}) =  \frac{2 g^4_{Z'}s \, (U^{\dagger}_{\mu i}U_{\mu j}-U^{\dagger}_{\tau i}U_{\tau j})^2}{3\pi [(s-m^2_{Z'})^2 + m^2_{Z'} \Gamma^2_{Z'}]},
\end{align}
where $U_{\alpha  i}$ is the Pontecorvo-Maki-Nakagawa-Sakata (PMNS) matrix, and where Greek (Latin) indices denote flavor (mass) eigenstates.

High-energy astrophysical neutrinos are thought to be produced almost entirely through the decay of charged pions, yielding an initial flavor ratio of $\nu_e:\nu_\mu:\nu_\tau =  1:2:0$. Through oscillations, such neutrinos evolve to possess an approximately equal proportion of flavors, $\nu_e:\nu_\mu:\nu_\tau \approx  1:1:1$. The cosmic neutrino background, acting as a target, is also approximately flavor universal (see, for example, Ref.~\cite{Mangano:2005cc}). 

For the case of $m_{Z'} \ll 2 m_{\mu}$, the $Z'$ will decay almost entirely into neutrino-antineutrino pairs of muon or tau flavor, with a total width that is given by
\begin{align}
\Gamma_{Z'} = \frac{g_{Z'}^2 m_{Z'}}{12 \pi}\,.
\label{eq:width1}
\end{align}

The neutrino spectrum that reaches Earth can be calculated by solving the following set of coupled differential equations~\cite{DiFranzo:2015qea}:
\begin{widetext}
\begin{align}
    -(1+z) \frac{H(z)}{c} \frac{\d \tilde{n}_{i}}{\d z} = J_{i}(E_{0},z) - \tilde{n}_{i} \sum_{j} \langle n_{\nu j}(z) \sigma_{ij}(E_{0},z) \rangle + P_{i} \int_{E_{0}}^{\infty} \d E' \sum_{j,k} \tilde{n}_{k} \left\langle n_{\nu j}(z) \frac{\d \sigma_{kj}}{\d E_{0}} (E',z)\right \rangle,
    \label{eq1}
\end{align}
\end{widetext}
where 
\begin{align}
    \tilde{n}_{i} \equiv& \frac{\d N_{i}}{\d E} (E_{0},z), \\[3pt]
     \quad P_{i} \equiv& \sum_{l} {\rm Br}(Z' \rightarrow \nu_{l}\nu_{i}), \nonumber
\end{align}
and $N_{i}$ is the comoving number density of neutrinos in the $i$-th mass eigenstate. Note that if the $Z'$ can only decay into neutrinos, then  $\sum_i P_i=1$ and the $P_i$'s are uniquely determined by the neutrino mass-mixing parameters. The Hubble rate in the redshift range of interest is given by
\begin{align}
H(z)\simeq H_0\sqrt{\Omega_\Lambda+\Omega_M(1+z)^{3}},
\end{align} 
where $H_0$ is the current value of the Hubble rate, $\Omega_M$ the matter density of the universe in units of the critical density, and $\Omega_\Lambda\simeq 1-\Omega_M$ is the dimensionless energy density associated with the cosmological constant. Throughout this paper, we adopt the best-fit cosmological parameters as reported by the Planck Collaboration~\cite{Planck:2018vyg}. The quantity, $E_0$, is the neutrino energy as measured at Earth. In absence of scattering, this is related to the energy at the source, $E$, according to $E=(1+z)E_0$. $n_{\nu j}(z)$ is the number density of neutrinos of mass eigenstate, $j$, in the cosmic neutrino background. The function $J_i(E_0, z)$ describes the spectrum and redshift distribution of the injected high-energy neutrinos. The second term on the right-hand side of Eq.~\eqref{eq1} accounts for the disappearance of high-energy neutrinos resulting from their scattering with the cosmic neutrino background, while the rightmost term describes the neutrinos that are produced in those scattering events. The differential scattering cross section entering the latter term is given by
\begin{align}
\frac{\d \sigma_{kj}}{\d E_0} (E',z) = \sigma_{kj}(E',z) f(E', E_0),
\end{align}
where $\sigma_{kj}$ is the total cross section and
\begin{align}
f(E',E_0) = \frac{3}{E'} \bigg[\bigg(\frac{E_0}{E'}\bigg)^2 + \bigg(1-\frac{E_0}{E'}\bigg)^2\bigg] \, \Theta(E'-E_0).
\end{align}
Note that $\int f(E',E_0) \d E_0 =2$ because each scattering event results in two outgoing high-energy neutrinos.

The thermally averaged quantity in the rightmost term of Eq.~\eqref{eq1} can be written as 
\begin{equation}
\label{thermalaverage}
    \langle n_{\nu j}(z) \sigma_{ij}(E_{0},z) \rangle  \equiv \int \frac{\d^3{\bf p}}{(2\pi)^3} \frac{\sigma_{ij}(E_0,z,{\bf p})}{e^{|{\bf p}|/T_0(1+z)}+1},
\end{equation}
where $\sigma_{ij}(E_0,z,{\bf p})$ can be found by evaluating Eq.~\eqref{sigma} with $s=2E_{\nu} \, [ (m^2_j + p^2)^{1/2} - |p| \cos \theta ]$, and $T_0\simeq 1.95\,$K is the temperature of the cosmic neutrino background at $z=0$. In the limit in which all of the neutrino masses are much larger than the effective temperature of the cosmic neutrino background, this quantity simplifies to $ \langle n_{\nu j}(z) \sigma_{ij}(E_{0},z) \rangle \simeq  n_{\nu j}(z) \sigma_{ij}(E_{0},z,0)$. 
In cases in which the mass of the lightest neutrino is not much greater than $T_0$, we calculate the thermal average as described in Eq.~\eqref{thermalaverage}.

In solving Eq.~\eqref{eq1}, we make use of neutrino number conservation, considering a series of redshift shells and evolving the neutrino energy spectrum at every redshift step. At each such step, we evolve the spectrum in each energy bin, adding the neutrinos that are injected from sources and subtracting those that scatter. Then, since we know the energy distribution of the outgoing neutrinos, we can appropriately redistribute the scattered neutrinos among the lower energy bins. For further details, see Ref.~\cite{DiFranzo:2015qea}.

\section{The impact of an MeV-scale gauge boson on the diffuse spectrum of high-energy  neutrinos}\label{fluximpact}

The IceCube neutrino observatory, which was completed in 2010, consists of an approximately cubic kilometer of Antarctic ice, with over 5000 digital optical modules distributed throughout its volume. This array of detectors is sensitive to the muon tracks and showers that are produced by high-energy neutrinos in and around the instrumented volume. IceCube has detected an approximately isotropic spectrum of diffuse astrophysical neutrinos, extending in energy between several TeV and several PeV~\cite{IceCube:2020acn,IceCube:2021rpz,IceCube:2013cdw,IceCube:2013low,IceCube:2014stg}. With the possible exceptions of the blazar TXS 0506+56~\cite{IceCube:2018dnn,IceCube:2018cha} and the nearby active galactic nucleus NGC 1068~\cite{IceCube:2022der}, these events have shown no significant correlation in either time or direction with any known astrophysical sources or classes of sources~\cite{IceCube:2019cia,IceCube:2016tpw,IceCube:2018omy,IceCube:2016ipa,Smith:2020oac,IceCube:2016qvd}. 

\begin{figure*}[t]
    \centering
    \includegraphics[width=0.65\textwidth]{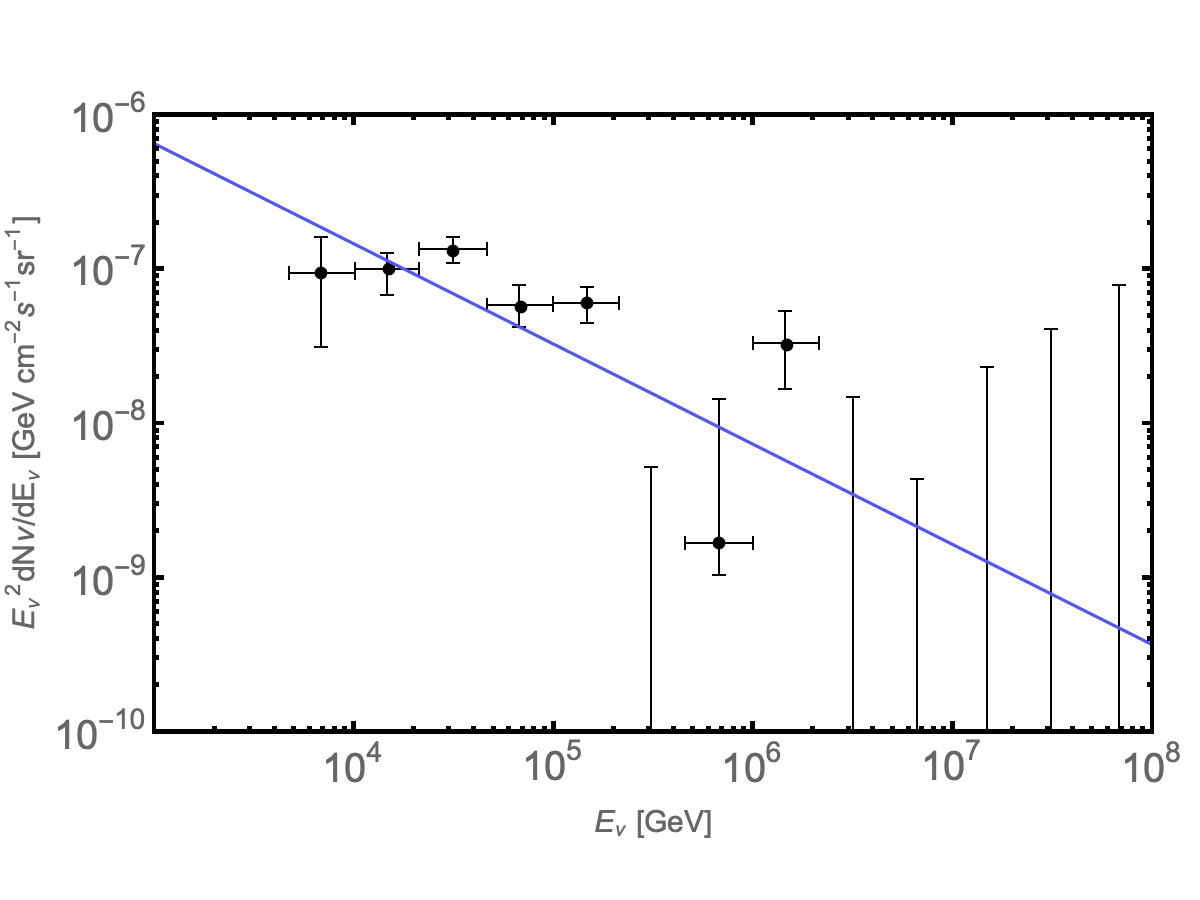}
    \caption{The spectrum of diffuse astrophysical neutrinos as reported by the IceCube Collaboration~\cite{IceCube:2020acn}, compared to the best-fit power-law.}
     \label{fig1}
\end{figure*}

In Fig.~\ref{fig1}, we show the diffuse neutrino spectrum as reported by the IceCube Collaboration~\cite{IceCube:2020acn}, and compare this to the best-fit power-law, which features an index of $\gamma = 2.65$. This power-law parameterization does not provide a particularly good fit to the measured spectrum, corresponding to $\chi^2=25.2$ (treating the reported errors as normally distributed). In particular, the measured spectrum shows signs of being flatter at the lowest measured energies, favoring $\gamma \sim 2.2$ below $\sim 100 \, {\rm TeV}$. Furthermore, the spectrum appears to be suppressed at energies between $E_{\nu} \sim 0.2-1 \, {\rm PeV}$. While this spectral feature could plausibly have something to do with the nature of the sources themselves, we will take this apparent ``dip'' in the spectrum to motivate models in which neutrinos in this energy range are significantly attenuated by the scattering induced by a new gauge boson. We will use this simple power-law fit to benchmark any improvement that might be provided by a model that includes a $Z'$.

\subsection{Results For a Simplified Source Distribution}\label{s2}

We now have all of the ingredients that we will need to calculate the impact of a $U(1)_{L_{\mu} - L_{\tau}}$ gauge boson on the spectrum of high-energy neutrinos observed at Earth. In this section, we compute the resulting spectrum for various choices of the source redshift distribution, and as a function of the injected spectral index, assuming a power-law form, $J(E_0,z) \propto E_0^{-\gamma}$.

The high-energy neutrino spectrum is resonantly attenuated when the total energy in the center-of-momentum frame is approximately equal to the mass of the gauge boson, $m_{Z'} \approx E_{\rm CM} \approx \sqrt{2 m_{\nu, i} E_{\nu}}$. This  produces an absorption feature in the observed spectrum at an energy given by
\begin{equation}
E_{\nu} \approx \frac{m_{Z'}^2}{2 m_{\nu, i} \, (1+z_{\rm abs}) }\approx 1\,{\rm PeV}\times \left(\frac{m_{Z'}}{10\,{\rm MeV}}\right)^2\left(\frac{0.05\,{\rm eV}}{m_{\nu, i}}\right) \left(\frac{1}{1+z_{\rm abs}}\right),
    \label{eq11}
\end{equation}
where $m_{\nu, i}$ is the mass of the $i$th neutrino species and $z_{\rm abs}$ is the redshift at which the scattering takes place. From this expression, we see that in order to obtain an absorption feature in the energy range favored by IceCube, we need to introduce a new gauge boson with $m_{Z'} \sim \mathcal{O}(10) \, {\rm MeV}$.

\begin{figure*}[t]
\centering
 \vspace{-1.7cm}
\includegraphics[width=0.48\textwidth]{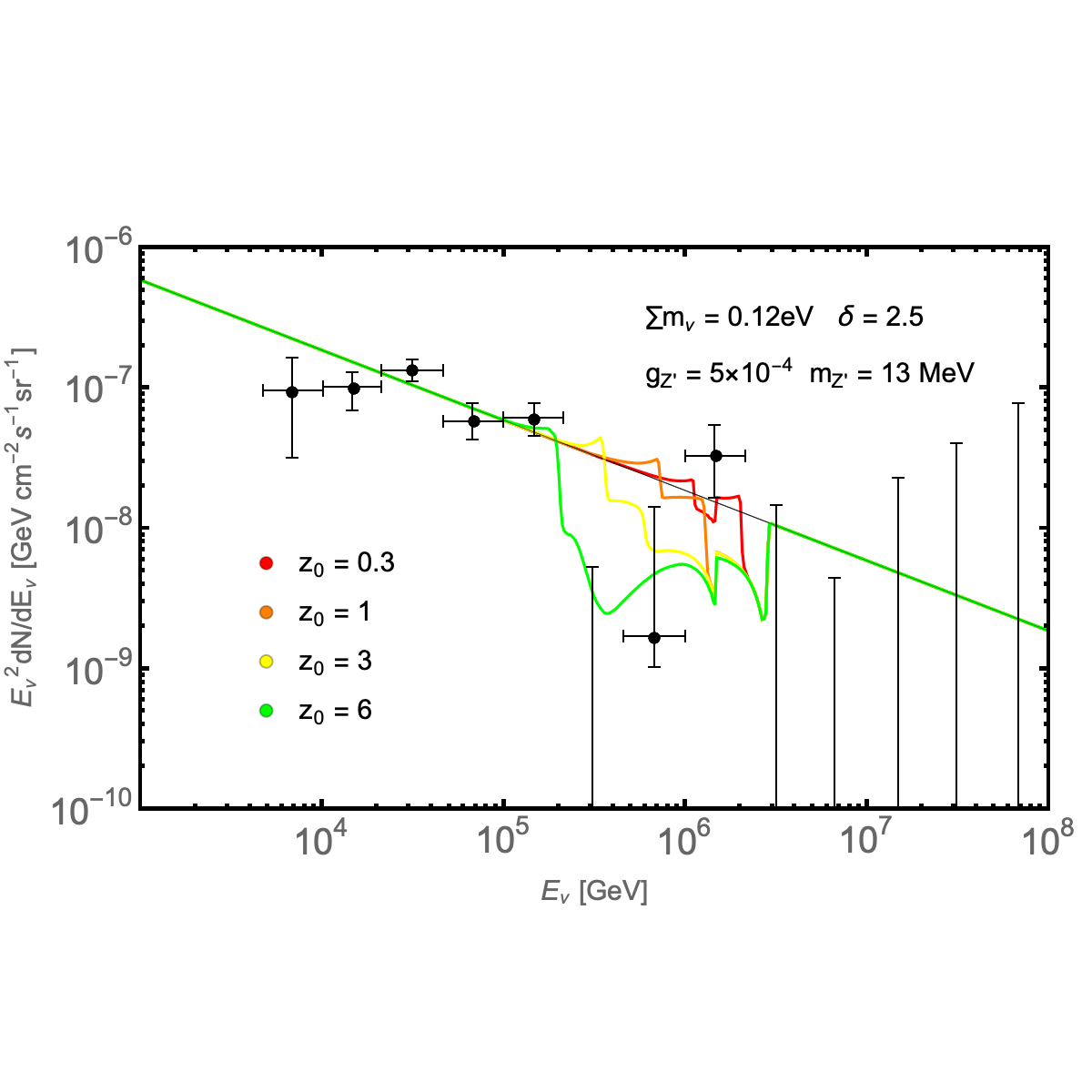}
\hspace{\fill}
\includegraphics[width=0.48\textwidth]{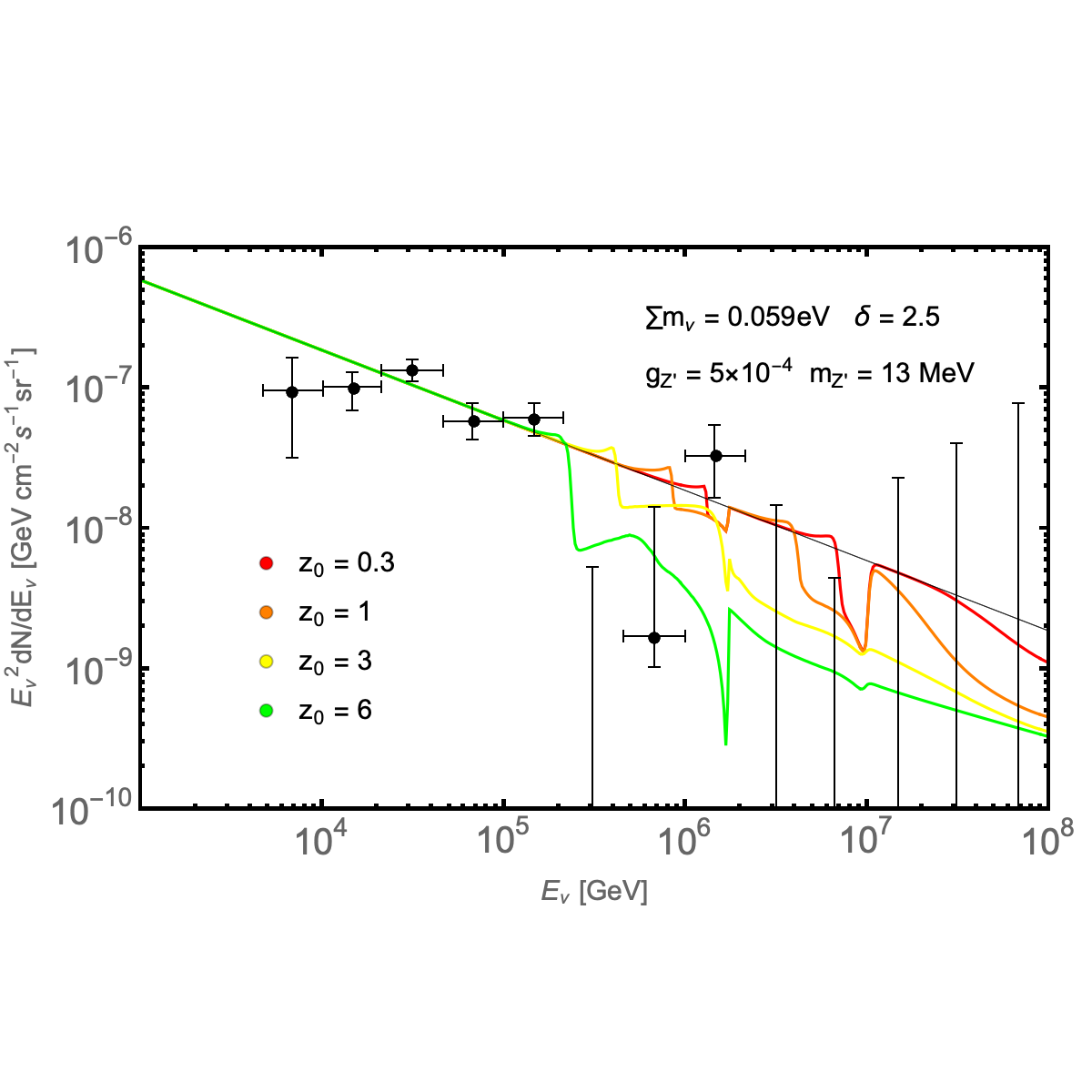}
 \vspace{-1cm}
\caption{The spectrum of high-energy neutrinos after including the effects of a $U(1)_{L_{\mu} - L_{\tau}}$ gauge boson with $m_{Z'}=13 \, {\rm MeV}$ and $g_{Z'}=5 \times 10^{-4}$ (chosen to accommodate the measured value of $g_{\mu}-2$). In each curve, the neutrinos are taken to originate from sources at a common redshift, $z_0$. In the left (right) frame, the sum of the neutrino masses is taken to be the maximum (minimum) allowed value. Here we have adopted the normal neutrino mass hierarchy, taken the injected spectral index to be $\gamma=2.5$, and fixed the normalization to obtain the best overall fit to the IceCube data.}
\label{fig2}
\end{figure*}

\begin{figure*}[!]
    \centering
    \vspace{-1.7cm}
    \includegraphics[width=0.48\textwidth]{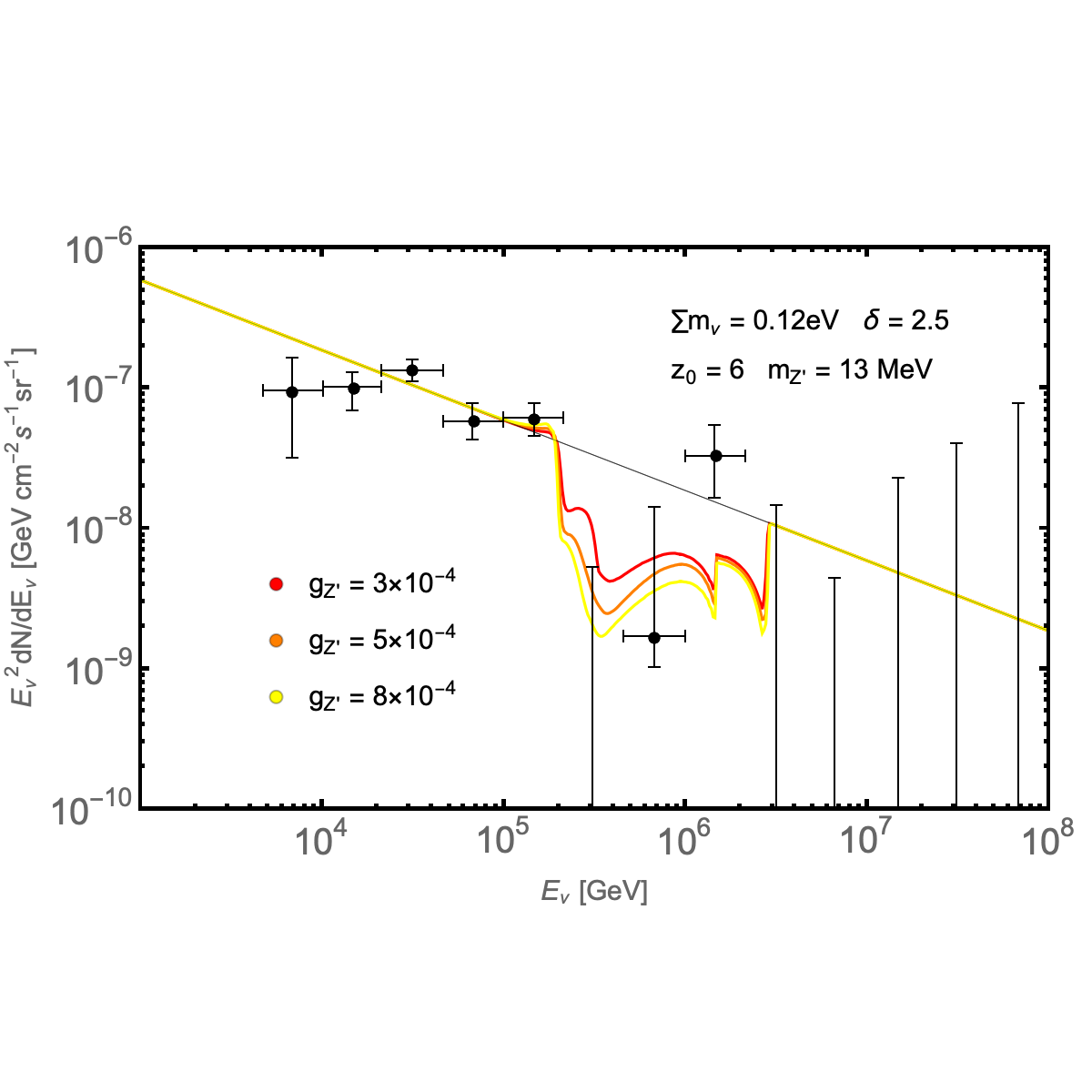}
        \includegraphics[width=0.48\textwidth]{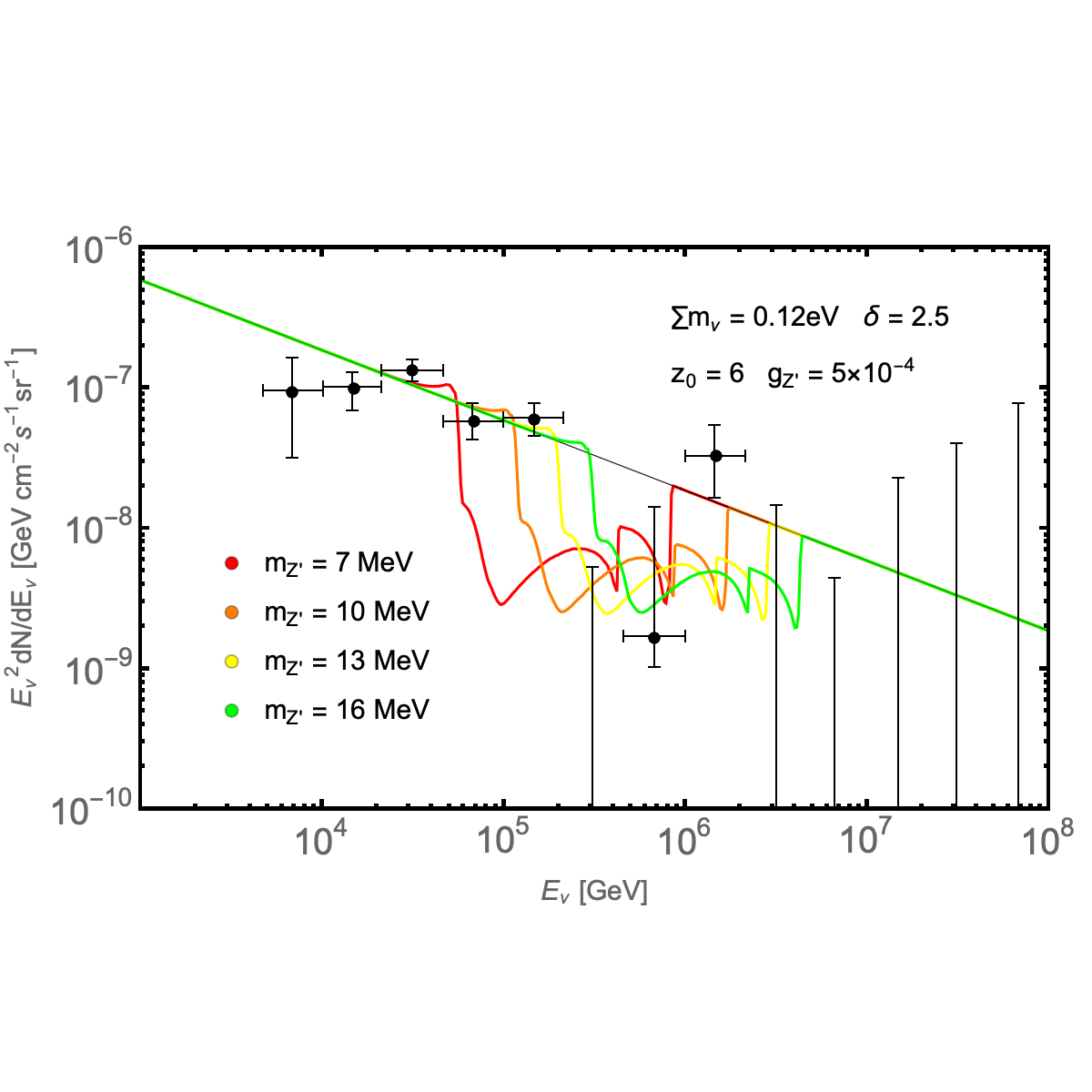}
            \vspace{-1cm}
    \caption{As in Fig.~\ref{fig2}, but fixing $z_0=6$ and varying either $g_{Z'}$ (left panel) or $m_{Z'}$ (right panel).}   
     \label{fig3}
\end{figure*}

As a first example, we will consider a toy model in which all of the high-energy neutrinos originate from sources at a common redshift, $z_0$: 
\begin{equation}
    J(E_0,z) \propto  E_0^{-\gamma} \, \delta(z-z_0).
    \label{zdirac}
\end{equation}

In Fig.~\ref{fig2}, we show the spectrum that results in this case for $m_{Z'}=13 \, {\rm MeV}$ and $g_{Z'}=5 \times 10^{-4}$ (chosen to accommodate the measured value of $g_{\mu}-2$), for four choices of $z_0$. In these frames, we have adopted the normal neutrino mass hierarchy and have set the sum of the three neutrino masses to either the maximum (0.12 eV) or minimum (0.059 eV) value allowed by oscillation data and cosmology~\cite{Planck:2018vyg}. We have further set the injected spectral index to $\gamma=2.5$ and fixed the normalization in each case in order to obtain the best overall fit to the IceCube data.

In the case shown in the left frame, the three neutrinos are approximately degenerate in mass, causing their corresponding absorption features to appear over approximately the same range of energies. In contrast, in the right frame, the lightest neutrino is massless, leading to absorption over a much wider range of $E_{\nu}$. Also notice that the overall magnitude of the resulting attenuation is larger and extends to lower energies for neutrinos that originate from higher redshift sources.  Lastly, in addition to the attenuation of this spectrum, one can identify in this figure small bump-like features which result from neutrinos being produced in $Z'$ mediated scattering events. In Fig.~\ref{fig3}, we show how these results change for different values of $g_{Z'}$ and $m_{Z'}$, for the case of $z_0=6$, the maximum value for the sum of the neutrino masses (0.12 eV), and again adopting the normal neutrino mass hierarchy.

\begin{figure*}[t]
    \centering
    \includegraphics[width=0.8\textwidth]{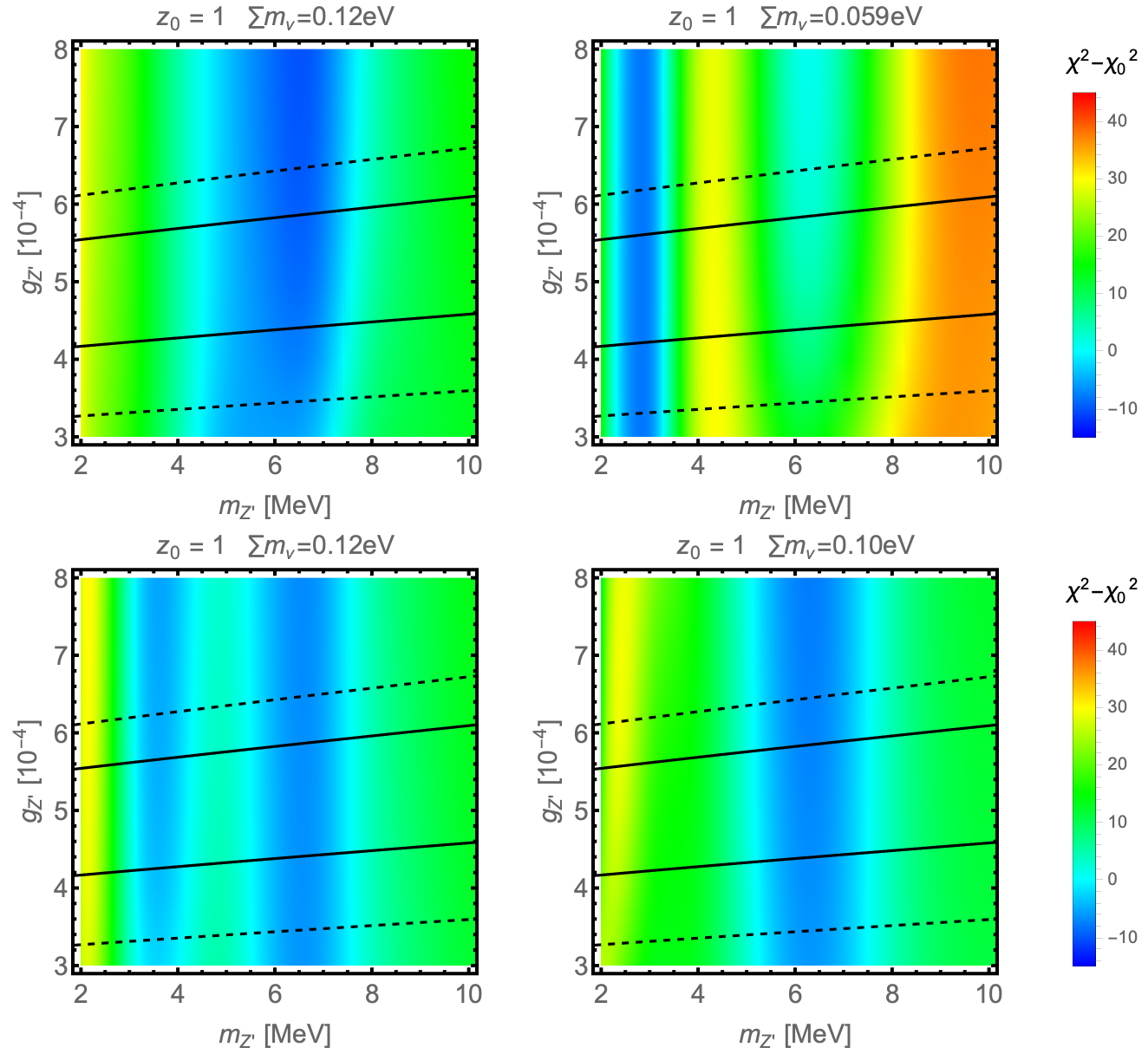}
    \caption{The improvement in the fit to the high-energy neutrino spectrum as a function of $m_{Z'}$ and $g_{Z'}$, for the case of $z_0=1$, and relative to the best-fit power-law without any new gauge boson. Results are shown for different values of the sum of the neutrino masses, and for the case of the normal (top frames) or inverted (bottom frames) mass hierarchy. In the regions between the solid (dashed) red lines, the $Z'$ can resolve the discrepancy between the predicted and measured values of the muon's anomalous magnetic moment to within 1$\sigma$ (2$\sigma$).}
    \label{fig5}
\end{figure*}

\begin{figure*}[htb]
\centering
 \vspace{-1.7cm}
\includegraphics[width=0.48\textwidth]{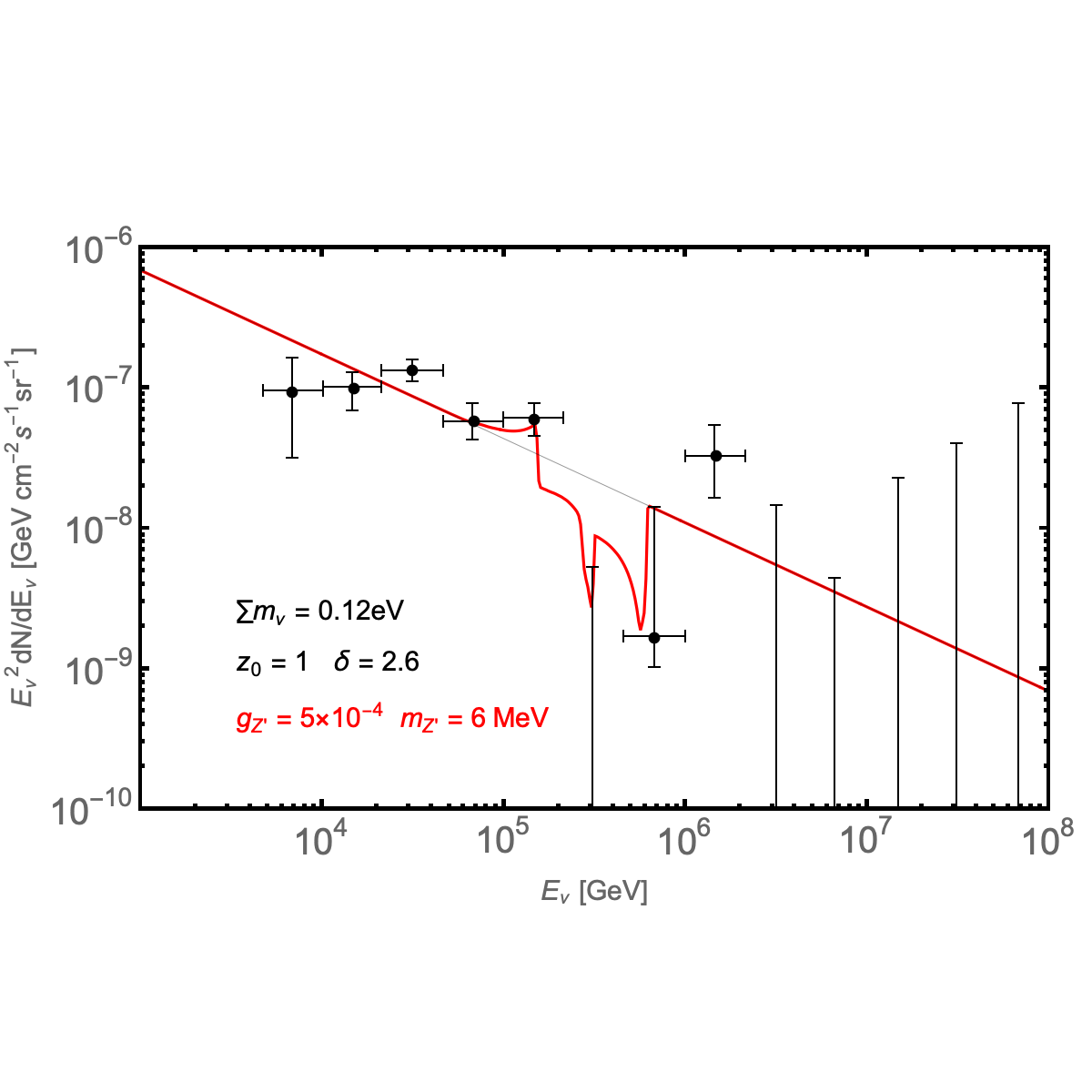}
\hspace{\fill}
    \includegraphics[width=0.48\textwidth]{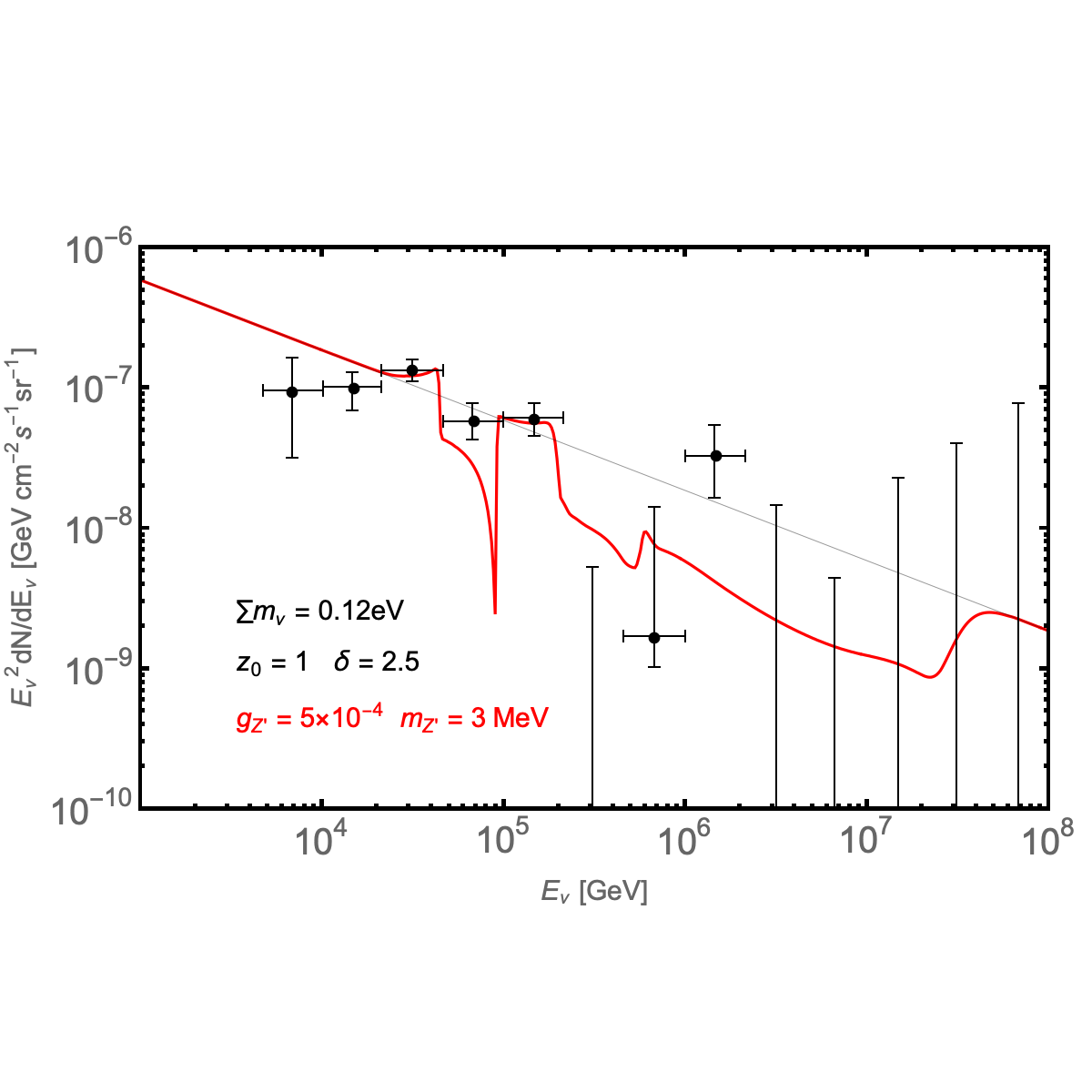}
     \vspace{-3cm}
   \\
    \includegraphics[width=0.48\textwidth]{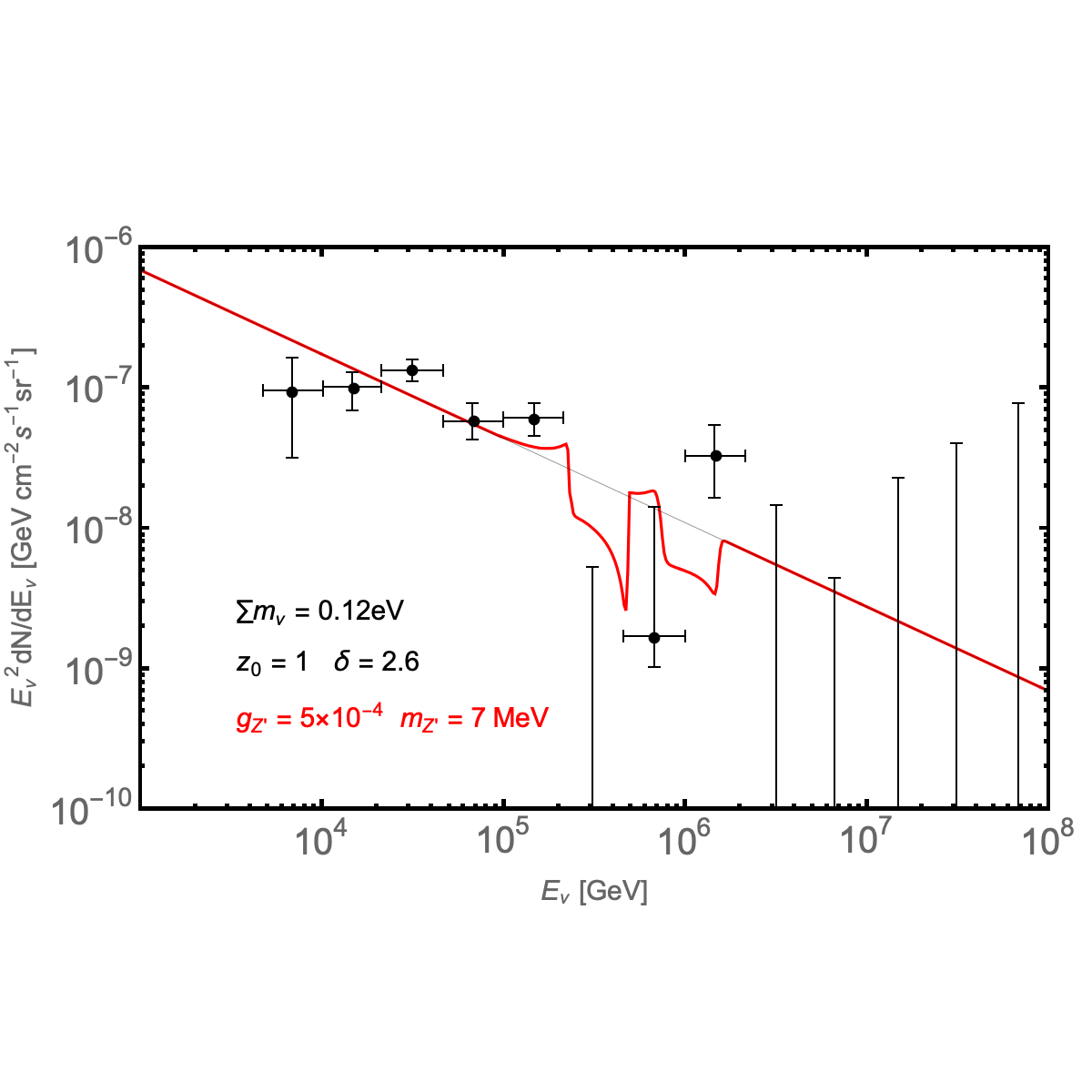}
    \hspace{\fill}
    \includegraphics[width=0.48\textwidth]{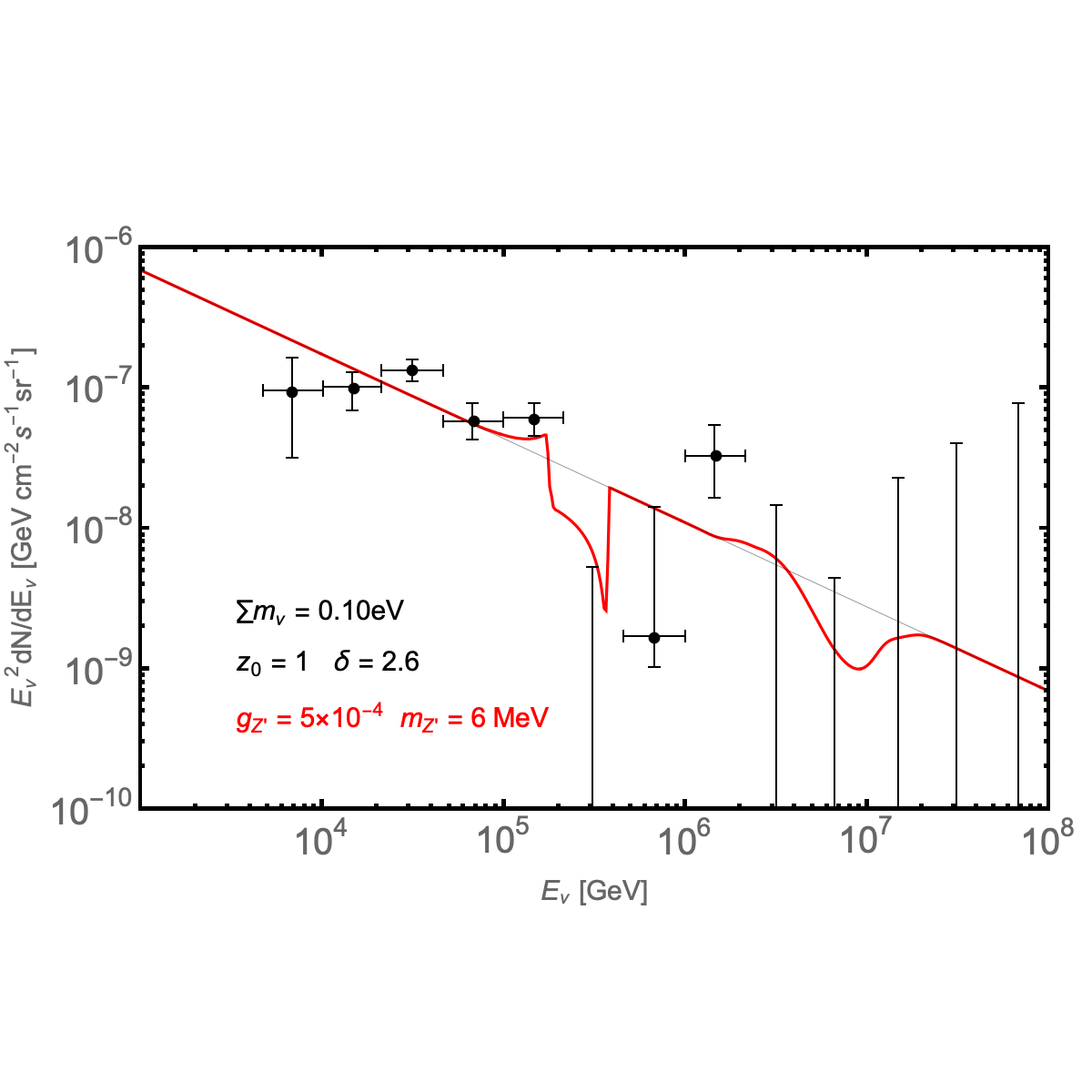}
    \vspace{-1cm}
\caption{The neutrino spectrum that is obtained for four selected sets of parameters. In the top (bottom) frames, the normal (inverted) mass hierarchy has been adopted.}
\label{fig6}
\end{figure*}

To evaluate the extent to which a new gauge boson could potentially improve the fit to the spectrum reported by the IceCube Collaboration, we have evaluated the value of the $\chi^2$ (again, treating the reported errors as normally distributed) to this data as a function of $m_{Z'}$, $g_{Z'}$, and $z_0$, considering different values for the sum of the neutrino masses, and for the normal or inverted mass hierarchy. We adopt the best-fit neutrino mixing parameters as reported in Ref.~\cite{de_Salas_2021}, and profile over the injected spectral index and normalization. In Fig.~\ref{fig5}, we compare the value of the $\chi^2$ obtained in this exercise (for the case of $z_0=1$) to the best-fit found without a new gauge boson, as shown in Fig.~\ref{fig1} and which yields $\chi_0^2=25.2$. In the case of the normal mass hierarchy and $\sum m_\nu \approx 0.12 \, {\rm eV}$, the quality of the fit can be improved substantially for $m_{Z'} \sim 4-8 \, {\rm MeV}$. In the case of the inverted hierarchy, significant improvement can be found for a similar range of masses.

In Fig.~\ref{fig6}, we show the spectrum obtained for four specific choices of parameters. In each of these cases, the attenuation caused by the new gauge boson significantly improves the quality of the fit to the spectrum reported by IceCube.
 The best fits are generally obtained for scenarios in which the sum of the neutrino masses is not too far below the maximum value allowed by cosmological considerations, $\sum m_{\nu} \sim 0.10-0.12 \, {\rm eV}$~\cite{Planck:2018vyg}, and for $m_{Z'}\sim 3-8 \, {\rm MeV}$. Such a small value of $m_{Z'}$ is in some tension with cosmological constraints, in particular through the resulting contribution to $N_{\rm eff}$~\cite{Escudero:2019gzq}. This can be relaxed, however, if the neutrinos originate predominantly from high-redshift sources, allowing good fits to be obtained for larger values of $m_{Z'}$. Also note that the precise value of the gauge coupling has only a modest impact on our results (at least within the range favored by $g_{\mu}-2$). This makes this probe complementary to laboratory constraints on such particles, such as that from the CCFR experiment~\cite{Altmannshofer:2014pba}.

\subsection{Results For Realistic Source Distributions}\label{s3}

Next, we will consider the impact of a new gauge boson on the spectrum of high-energy neutrinos, adopting some examples of well-motivated source distributions. In particular, we will perform these calculations using source distributions which trace the observed population of BL Lacertae objects (BL Lacs)~\cite{Ajello:2013lka}, or the observed rate of star formation~\cite{Yuksel:2008cu}.

\begin{figure}[t]
\centering
\includegraphics[width=0.75\textwidth]{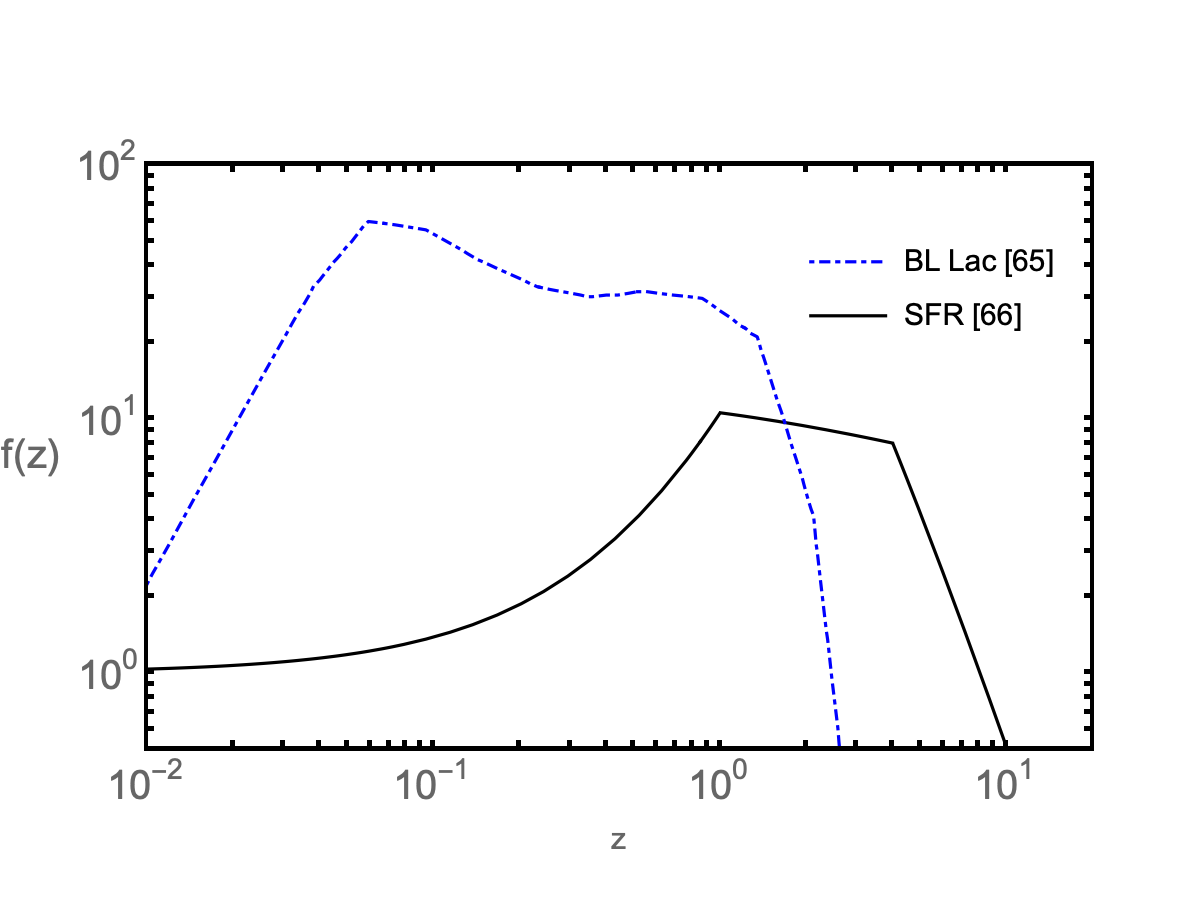}
\caption{The BL Lac redshift distribution, $f_{\rm BL}$, as derived in Ref.~\cite{Ajello:2013lka} (dot dashed blue), and the star formation rate redshift distribution, $f_{\rm SFR}$, as derived in Ref.~\cite{Yuksel:2008cu} (black).}
\label{fig7}
\end{figure}

The redshift distribution of BL Lacs can be expressed as follows~\cite{Ajello:2013lka}:
\begin{equation}
    J_{\rm BL}(E_{\nu},z) \propto E^{-\gamma}_{\nu} \, (1+z)^{3} f_{\rm BL}(z),
\end{equation}

where the function, $f_{\rm BL}$, is shown in Fig.~\ref{fig7}.

For sources that trace the star formation rate, we adopt the following parameterization~\cite{Yuksel:2008cu}:

\begin{equation}
    J_{\rm SFR}(E_{\nu},z) \propto E^{-\gamma}_{\nu} \, (1+z)^{3} f_{\rm SFR}(z),
\end{equation}

where

\begin{equation}
    f_{\rm SFR} \propto \begin{cases}
            (1+z)^{3.4}, \quad \,\,\,\,\,\,\,\,\,\,\,\,\,\,\,\,\,\,\,\,\,\,\,\,\,\,\,\,\,\,\,\,\,\, z \leq 1 \\
            2^{3.7} \, (1+z)^{-0.3}, \quad \,\,\,\,\,\,\,\,\,1 < z < 4 \\
            2^{3.7} \,5^{3.2} \, (1+z)^{-3.5}, \,\,\,\,\,\,\,\,\,\,\,\,\,\,\, z \geq 4. \\
            \end{cases}
\end{equation}

While BL Lacs are largely found at relatively low redshifts, $z \lsim 1-2$, the star formation rate extends to much higher values of $z$ (see Fig.~\ref{fig7}).

\begin{figure*}[htb]
    \centering
    \includegraphics[width=0.8\textwidth]{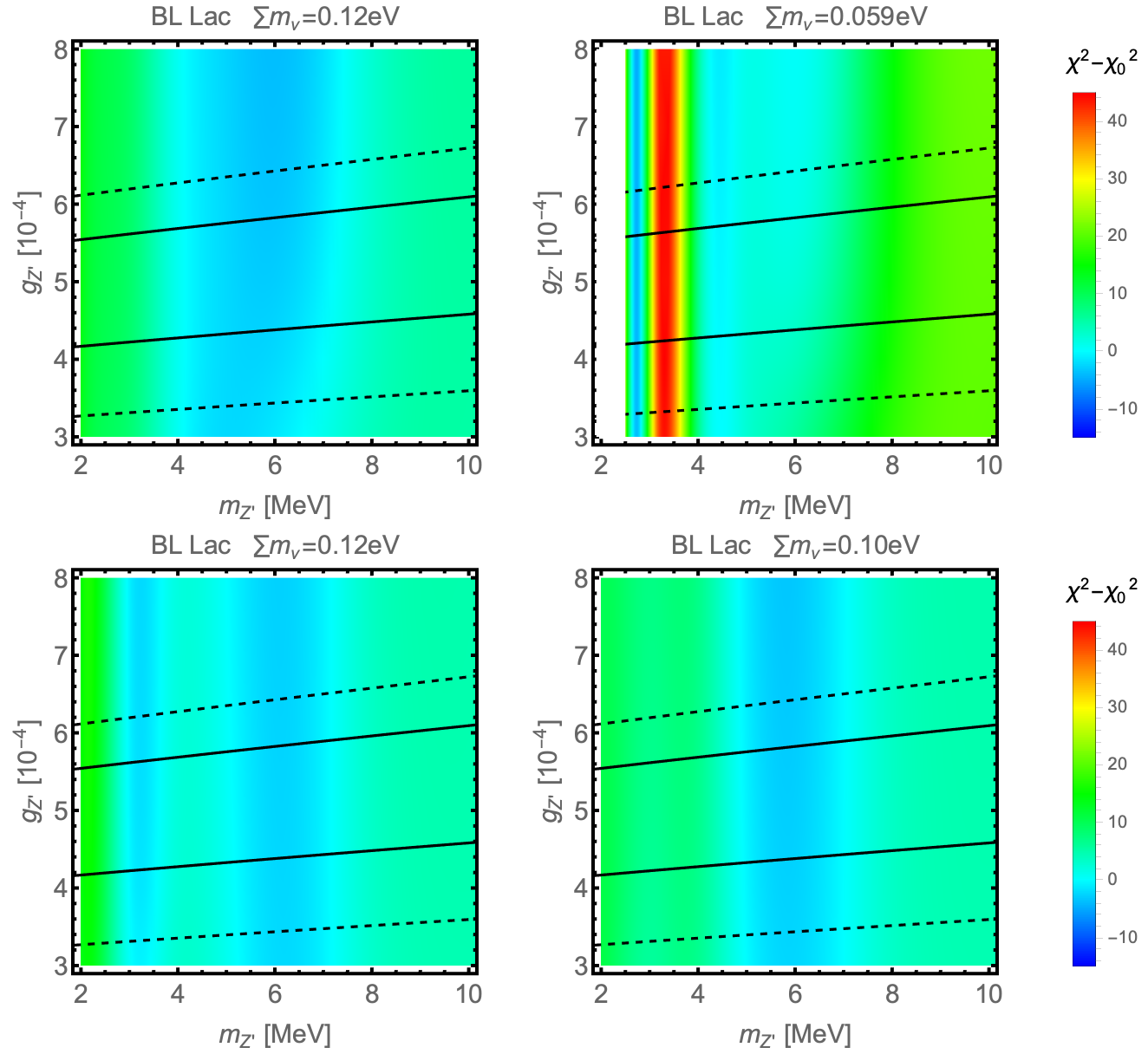} \\
    \vspace{-1cm}
 \includegraphics[width=0.48\textwidth]{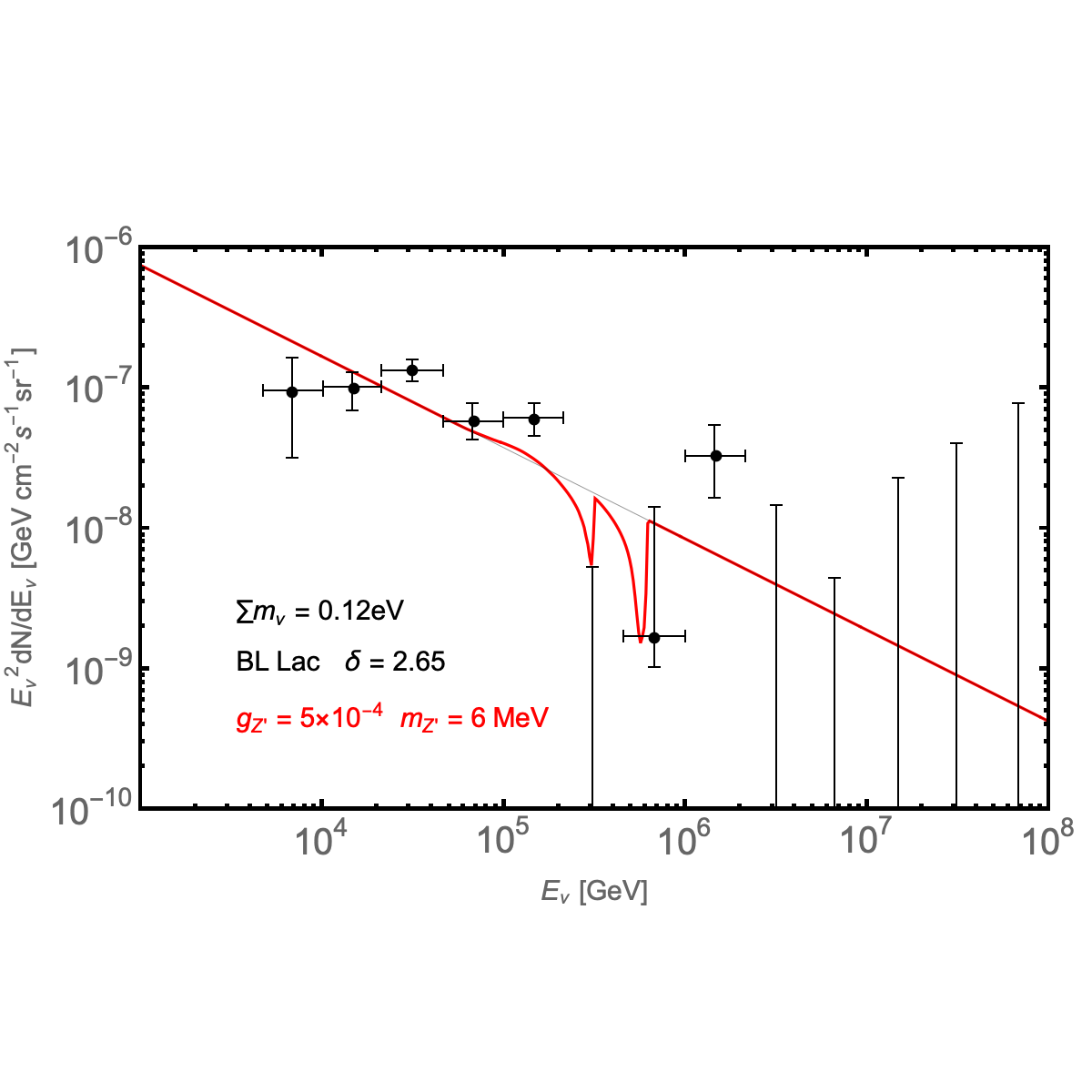}
\hspace{\fill}
\includegraphics[width=0.48\textwidth]{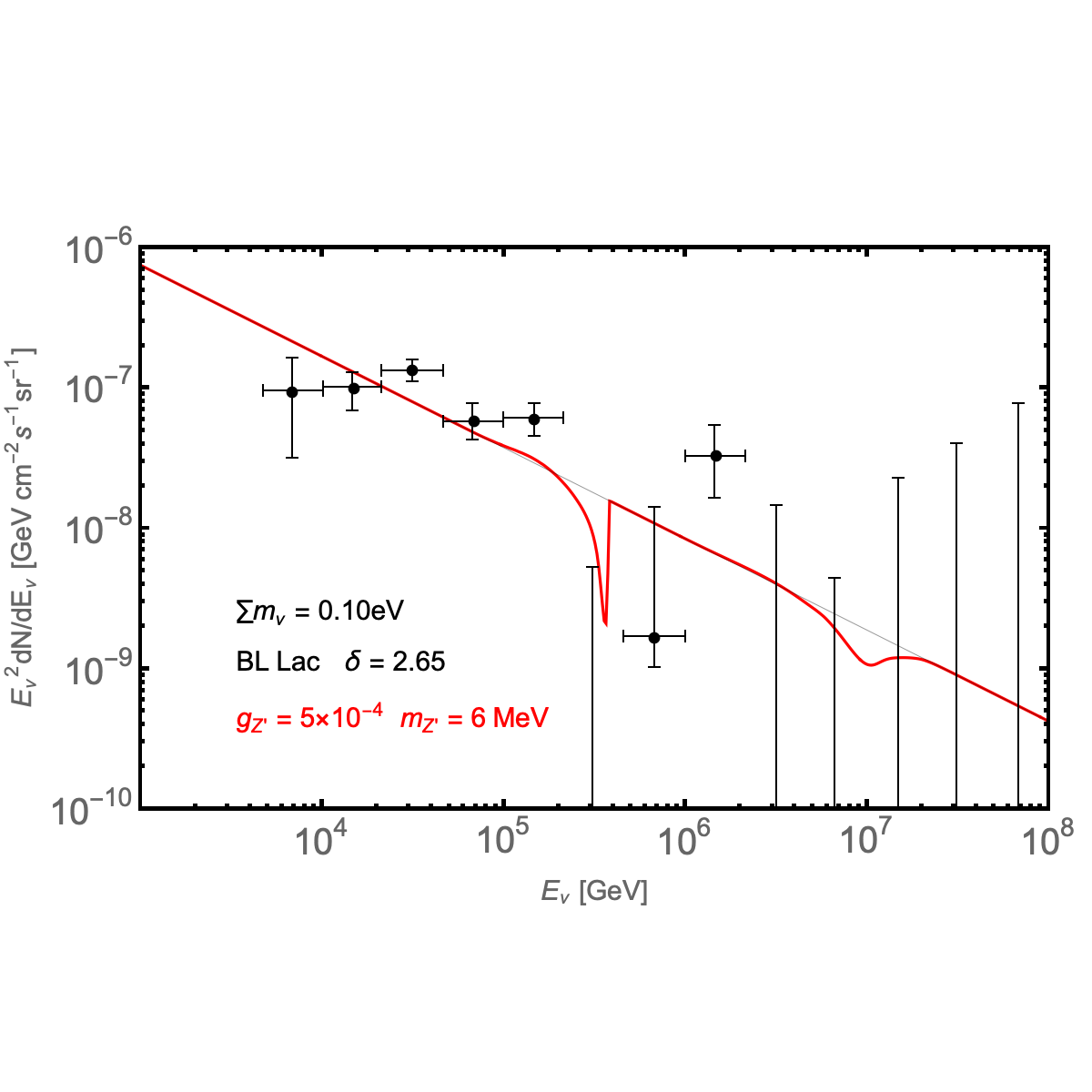}
\vspace{-1cm}
    \caption{As in Figs.~\ref{fig5} and~\ref{fig6}, but for a redshift distribution of high-energy neutrino sources that traces the observed distribution of BL Lacs. In the upper left, upper right, and lower left frames, the normal mass hierarchy has been adopted, while in the other frames we have used the inverted hierarchy.}
    \label{fig8}
\end{figure*}

\begin{figure*}[htb]
    \centering
    \includegraphics[width=0.8\textwidth]{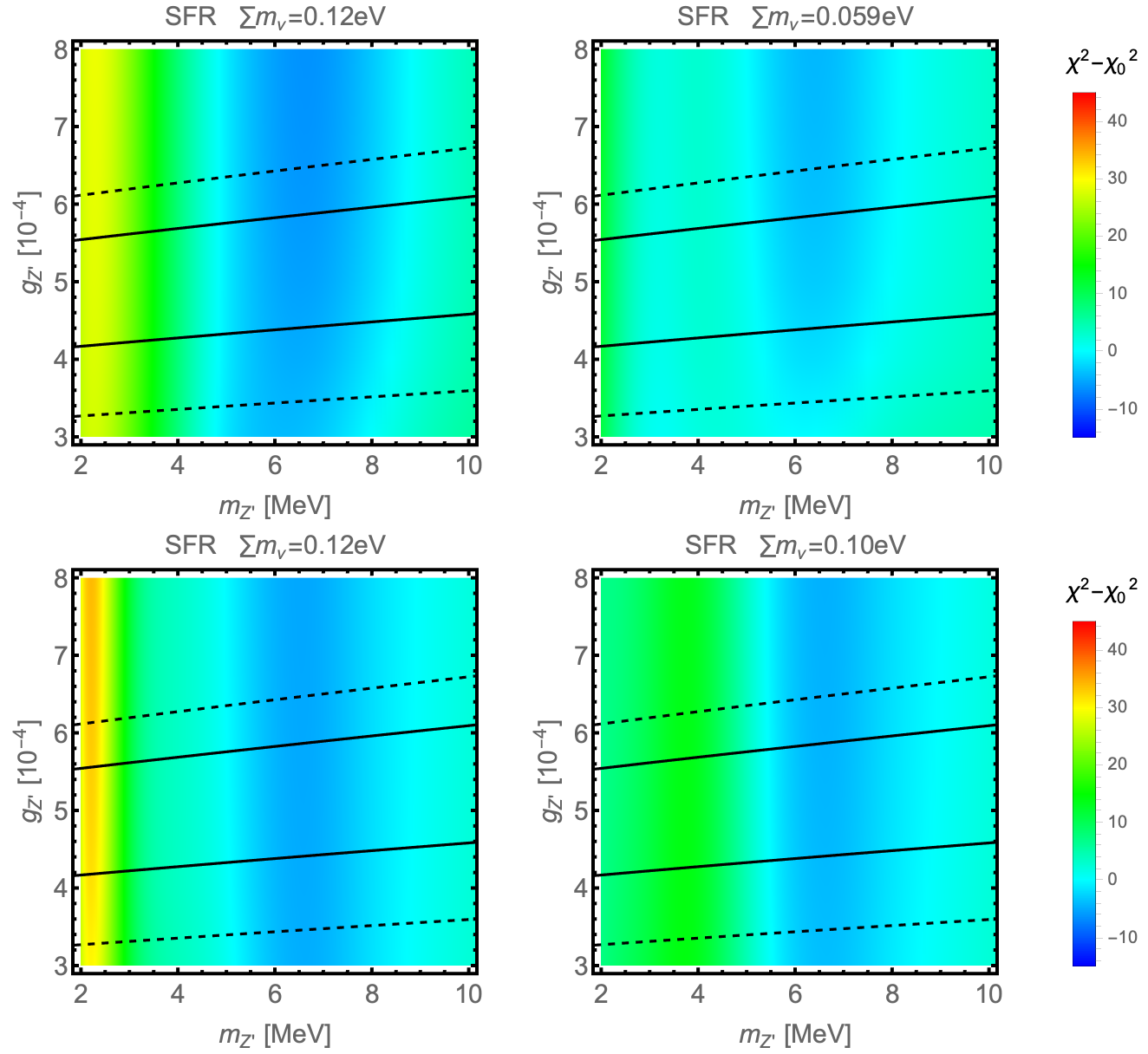} \\
    \vspace{-1cm}
 \includegraphics[width=0.48\textwidth]{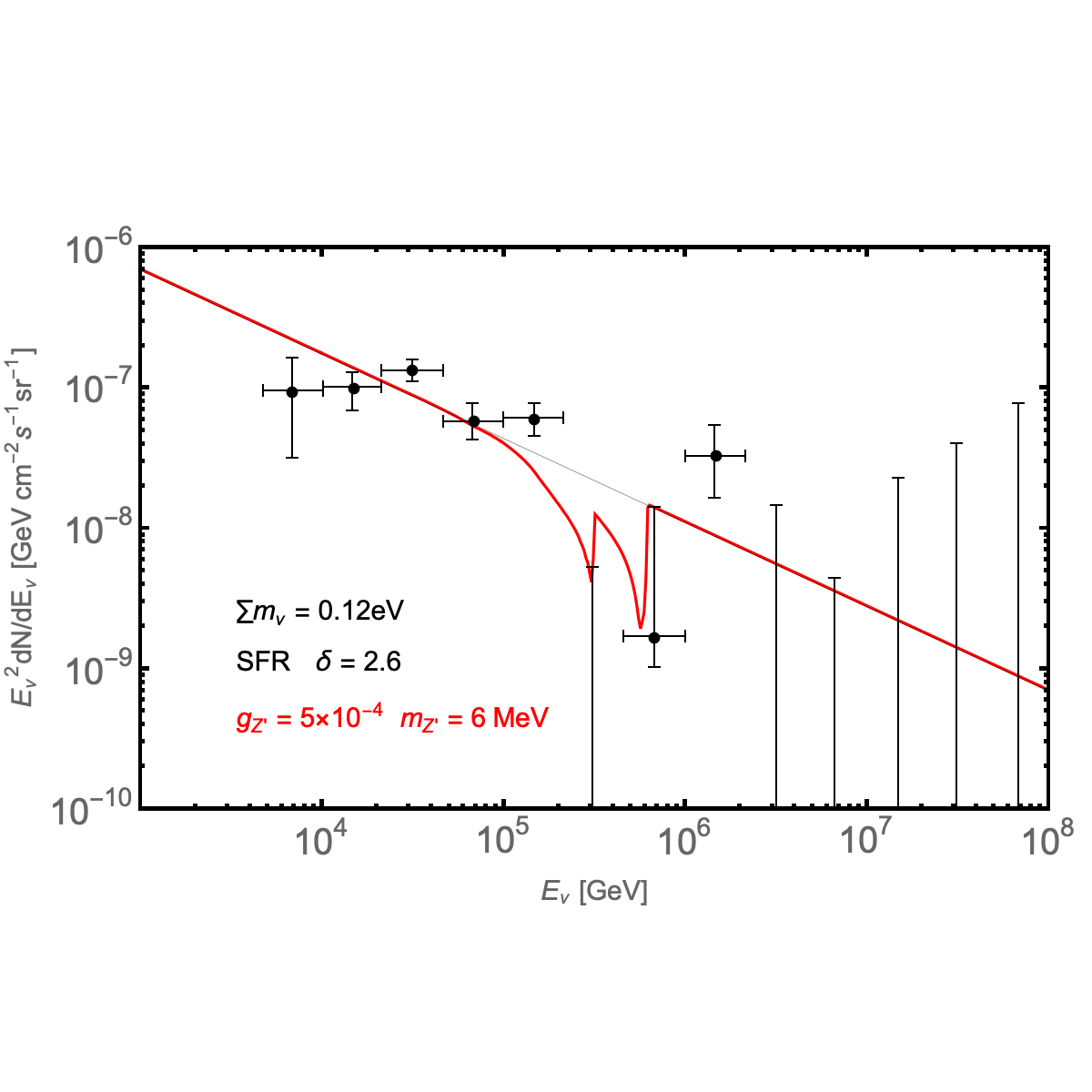}
\hspace{\fill}
\includegraphics[width=0.48\textwidth]{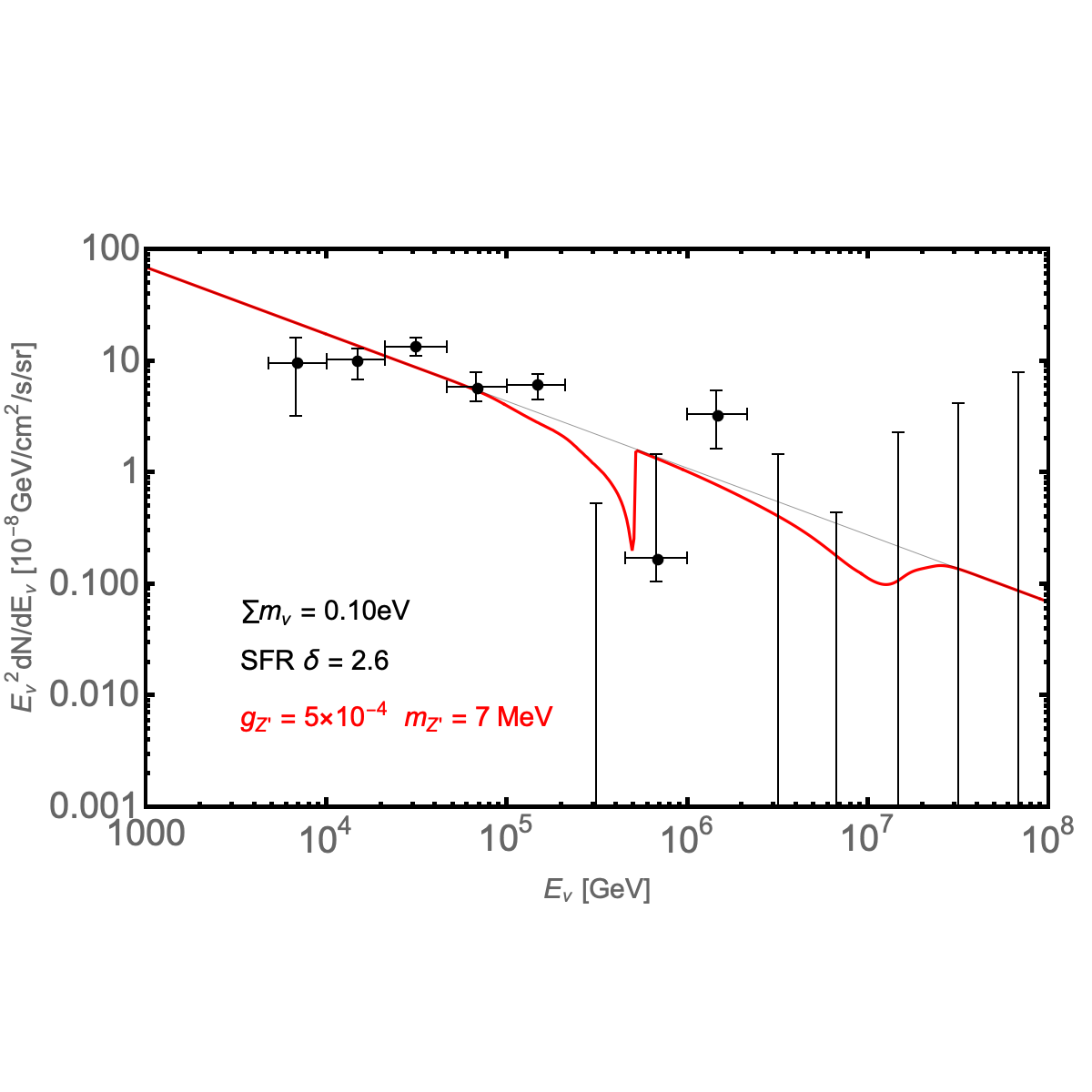}
\vspace{-1cm}
    \caption{As in Figs.~\ref{fig5},~\ref{fig6} and~\ref{fig8}, but for a redshift distribution of high-energy neutrino sources that traces the observed star formation rate. In the upper left, upper right, and lower left frames, the normal mass hierarchy has been adopted, while in the other frames we have used the inverted hierarchy.}
    \label{fig10}
\end{figure*}

The results obtained for the case of a BL Lac redshift distribution are shown in Fig.~\ref{fig8}. These results are similar to those found for the $z_0=1$ case, although with somewhat less overall attenuation. For the case of the normal hierarchy and a sum of neutrino masses near the lowest allowed value, IceCube data can be used to exclude a narrow band of parameter space around $m_{Z'} \sim 3-4 \, {\rm MeV}$, in which a distinctive attenuation feature would have been generated at $E_{\nu} \sim m^2_{Z'}(1+z_{\rm abs})/(2m_{\nu}) \sim 30-100 \, {\rm TeV}$. In the same frame of Fig.~\ref{fig8}, we see that larger values of the gauge boson mass, $m_{Z'} \gsim 7 \, {\rm MeV}$, are also significantly disfavored by the fit. This conclusion, however, relies on the assumption of a power-law injection spectrum, making robust conclusions difficult to draw at this time. In scenarios in which the lightest neutrino is not nearly massless, the spectrum reported by IceCube favors the presence of a new gauge boson with a mass in the range of $m_{Z'}\sim 5-7\,$MeV. For values of the gauge coupling that can explain the measured value of $g_{\mu}-2$, the presence of such a particle can improve the fit by up to $\Delta \chi^2 \sim 5$. Representative examples of such scenarios scenarios are shown in the bottom frames of Fig.~\ref{fig8}.

The redshift distribution of BL Lacs peaks at relatively low redshifts, limiting the degree of attenuation that is induced. If we instead consider sources of high-energy neutrinos that are distributed according to the star formation rate, the resulting spectral feature can be more pronounced, and can appear in the energy range favored by IceCube for larger values of $m_{Z'}$ (resulting in less tension with cosmological constraints~\cite{Escudero:2019gzq}). In Fig.~\ref{fig10}, we show the results that we obtain in this interesting case.

\begin{figure*}[!htb]
    \centering
    \includegraphics[width=0.8\textwidth]{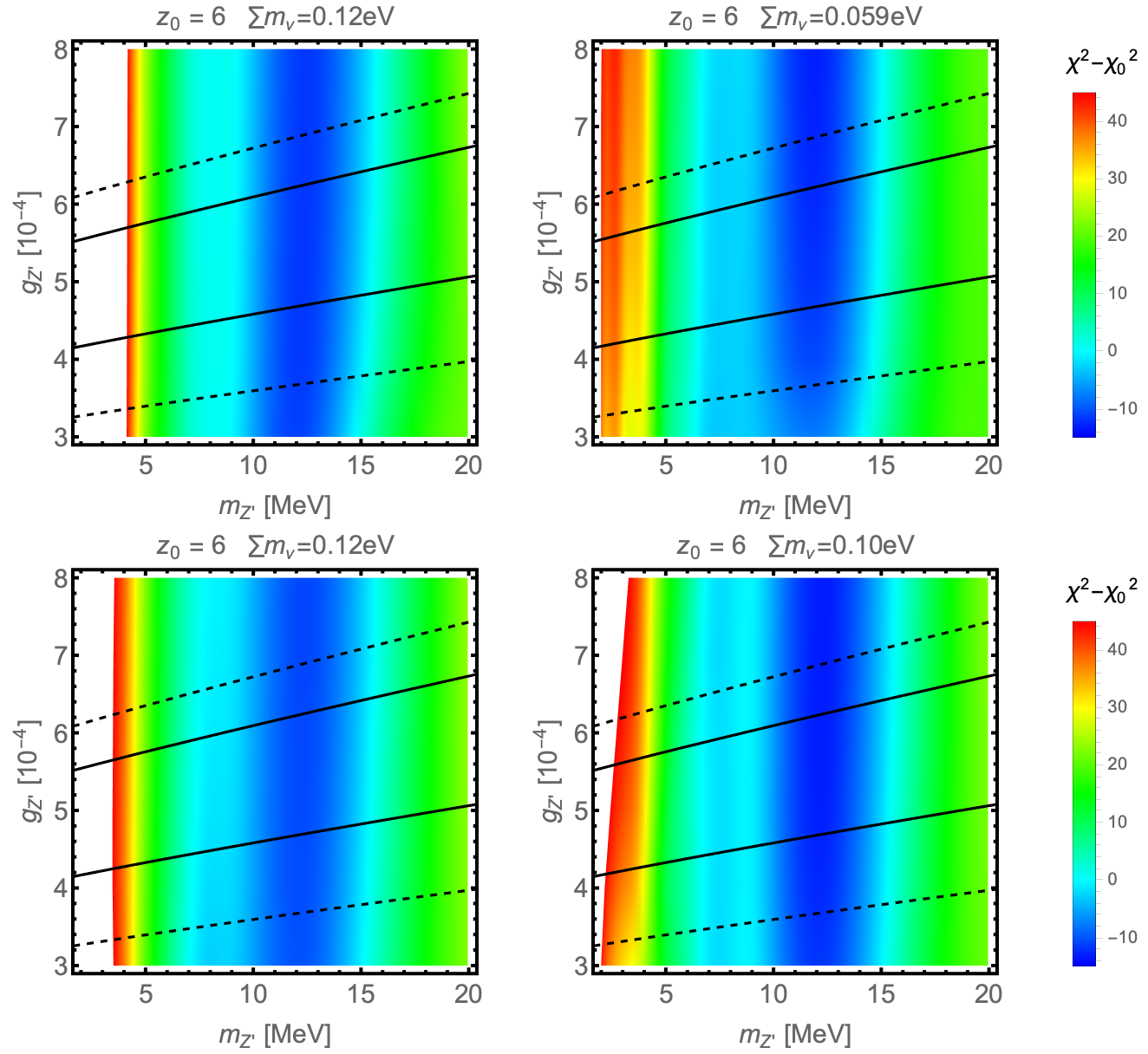}
    \caption{As in Figs.~\ref{fig5},~\ref{fig8} and~\ref{fig10}, but for high-energy neutrino sources that are all located at $z=6$. In the top (bottom) frames, the normal (inverted) mass hierarchy has been adopted.}
    \label{fig12}
\end{figure*}

As a final source distribution, we consider an example in which all of the neutrinos observed by IceCube originate from high-redshift sources, at $z=6$. While this is not a particularly realistic case, it can be taken to represent a class of scenarios in which the distribution of such sources peaks at high redshifts. The quality of our fits in this case are shown in Fig.~\ref{fig12}. With this redshift distribution, we can obtain fits which improve over the unattenuated power-law by $\Delta \chi^2 \sim 10-15$, and with values of $m_{Z'}$ that are large enough to be consistent with cosmological constraints~\cite{Escudero:2019gzq}.

\section{Dark Sector Models}\label{dark}

Up to this point, we have considered the impact of a new gauge boson without introducing any other additional particle content. It is possible, however, that such a gauge boson could also couple to light states that do not carry any Standard Model charges. In such a scenario, the scattering of high-energy neutrinos with the cosmic neutrino background could additionally result in the production of light dark sector states.

For gauge boson couplings to a light dark sector fermion of the form ${\cal L}_{\rm int} \supset g_{\chi}Z'_\mu \bar\chi\gamma^\mu\chi$, we can modify Eq.~\eqref{eq1} to obtain the cross section for dark sector particle production:
\begin{align}
\sigma(\nu_i  \bar{\nu}_j \rightarrow \chi \bar{\chi}) \simeq  \frac{g_\chi^2 g^2_{Z'}s \, (U^{\dagger}_{\mu i}U_{\mu j}-U^{\dagger}_{\tau i}U_{\tau j})^2}{3\pi [(s-m^2_{Z'})^2 + m^2_{Z'} \Gamma^2_{Z'}]},
\end{align}
where $\Gamma_{Z'}$ is equal to the sum of the width to $\nu \bar{\nu}$, as given in Eq.~\ref{eq:width1}, and the following contribution to $\chi \bar{\chi}$:
\begin{align}
\Gamma_{Z' \rightarrow \chi \bar{\chi}}=\frac{g_\chi^2m_{Z'}}{12\pi} \left(1+\frac{2m_\chi^2}{m_{Z'}^2}\right)\sqrt{1-\frac{4m_\chi^2}{m_{Z'}^2}} \simeq 
\frac{g_\chi^2m_{Z'}}{12\pi}.
\end{align}

Note that the regeneration feature in the spectrum only receives a contribution from $Z' \rightarrow \nu \bar{\nu}$, resulting in a rescaling of $P_i\to r\,P_i$, where $r=2\Gamma_{\nu\bar{\nu}}/(2\Gamma_{\nu\bar{\nu}}+\Gamma_{\chi\bar{\chi}})$. In the limit of $g_{Z'}^{2} \gg g_{\chi}^{2}$, we recover the same results as presented in the previous sections. In the opposite case of $g_{\chi}^{2} \gg g_{Z'}^{2}$ and $m_\chi\ll m_{Z'}$, the $Z'$ will decay almost entirely to dark sector particles, enhancing the neutrino scattering rate while also suppressing any neutrino regeneration features.

\begin{figure}[t]
\centering
 \vspace{-0.2cm}
\includegraphics[width=0.5\textwidth]{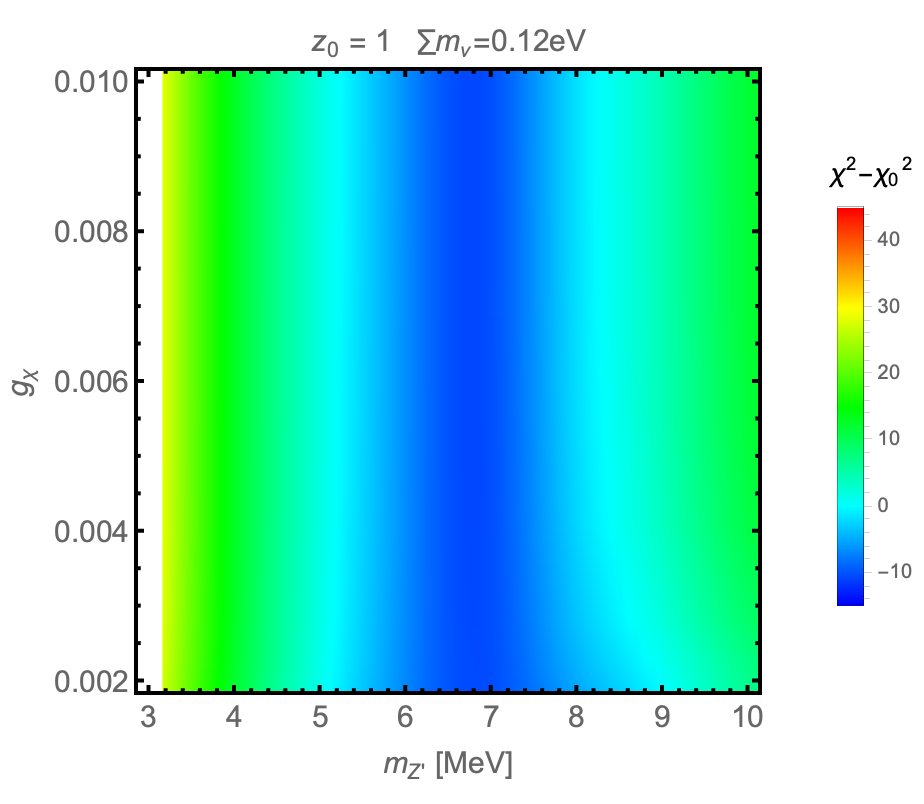}
\hspace{\fill}
\includegraphics[width=0.44\textwidth]{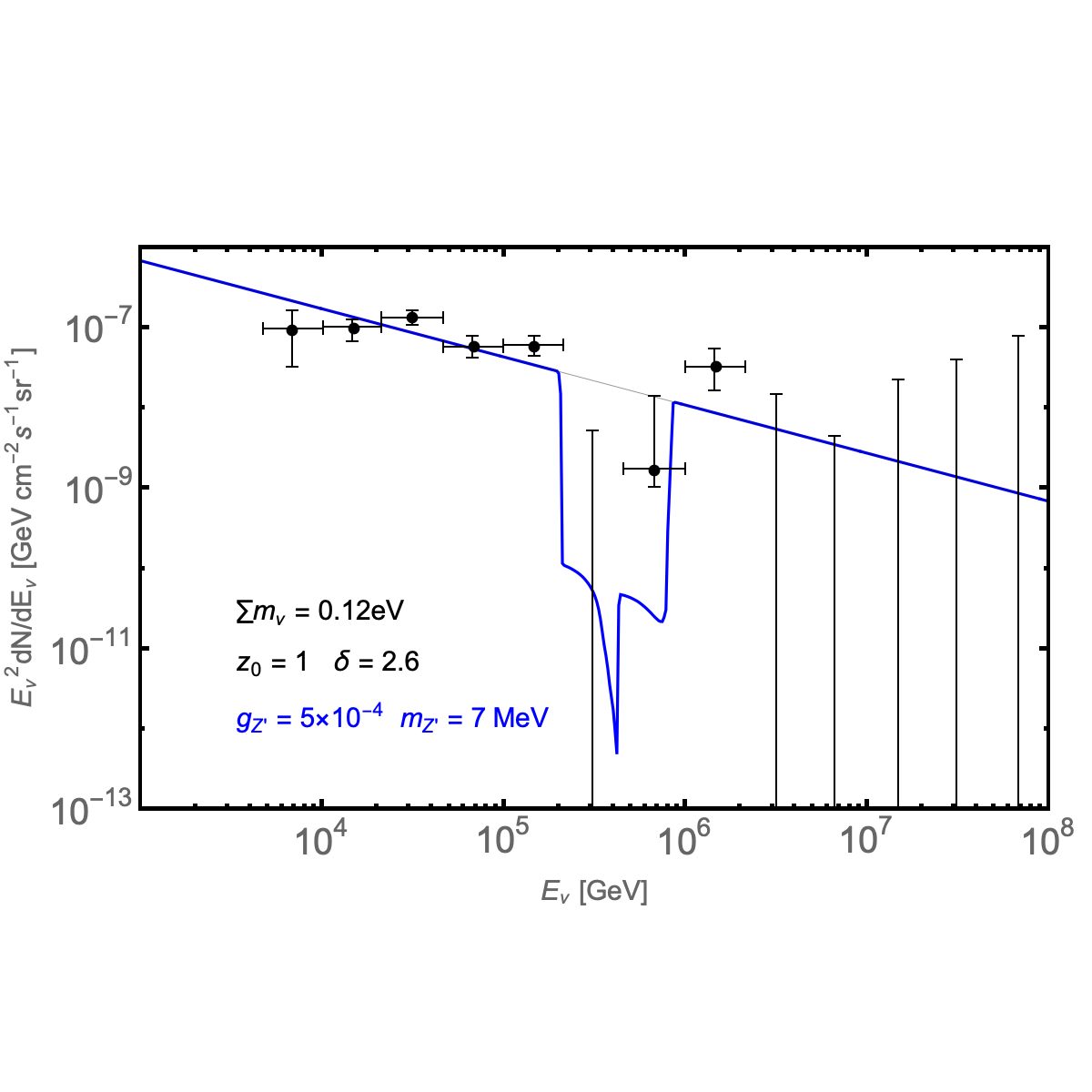}
 \vspace{0cm}
\caption{As in previous figures, but allowing for the decays of the $Z'$ into dark sector fermions, $Z' \rightarrow \chi \bar{\chi}$. Here we have taken the neutrinos to originate from sources at a common redshift of $z=1$, adopted $g_{Z'}=5\times 10^{-4}$, $g_{\chi} = 7 \times 10^{-3}$, and the normal neutrino mass hierarchy.}
\label{fig13}
\end{figure}

\begin{figure}[t]
\centering
 \vspace{-0.2cm}
\includegraphics[width=0.5\textwidth]{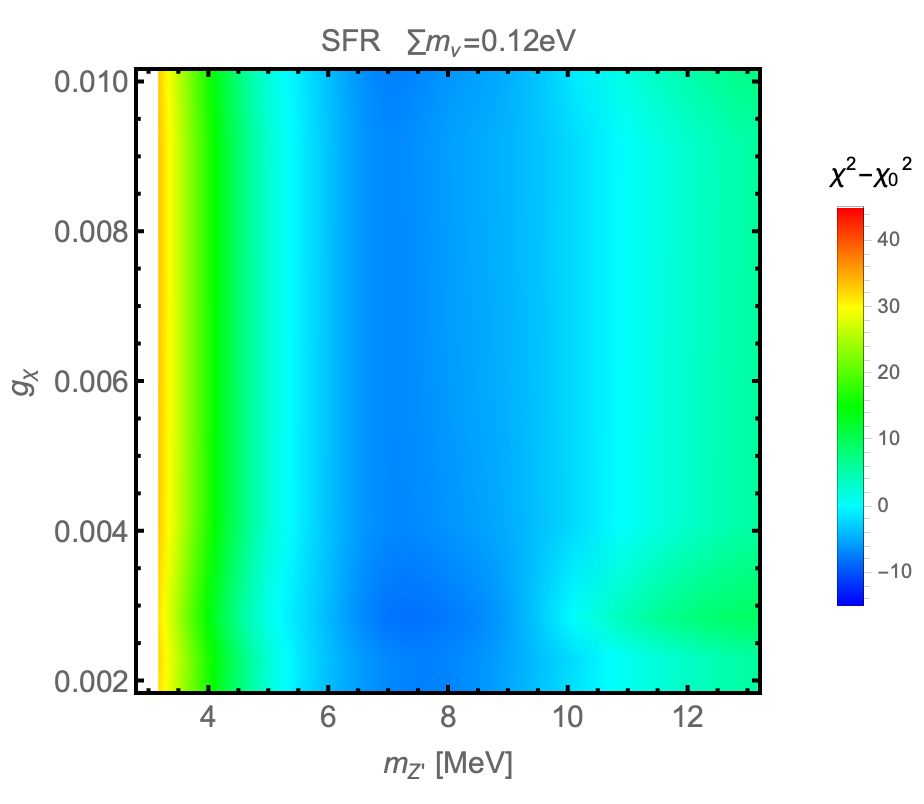}
\hspace{\fill}
\includegraphics[width=0.44\textwidth]{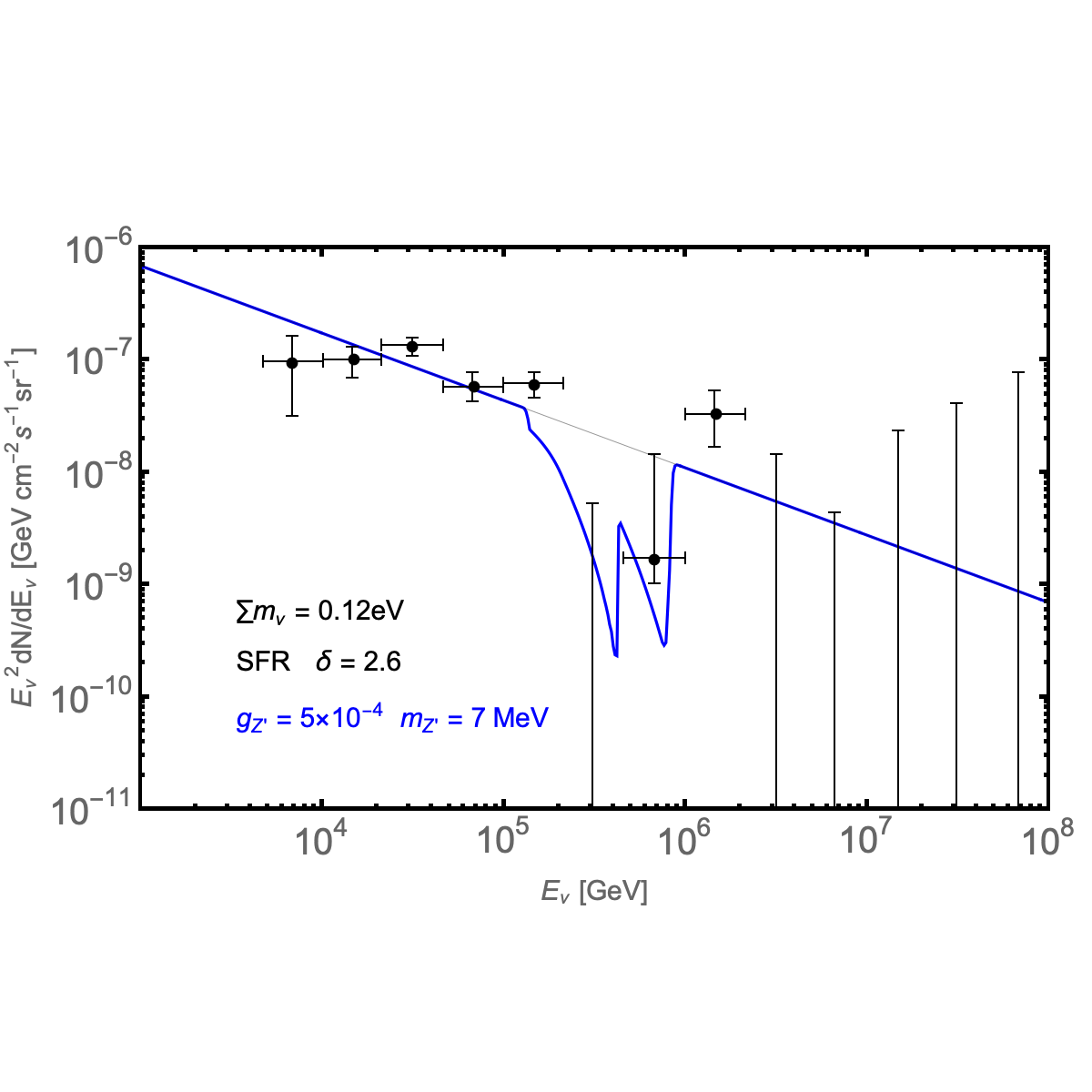}
 \vspace{0cm}
\caption{As in previous figures, but allowing for the decays of the $Z'$ into dark sector fermions, $Z' \rightarrow \chi \bar{\chi}$. Here we have taken the neutrinos to originate from sources distributed according to the star formation rate, adopted $g_{Z'}=5\times 10^{-4}$, $g_{\chi} = 3 \times 10^{-3}$, and the normal neutrino mass hierarchy.}
\label{fig14}
\end{figure}

In Figs.~\ref{fig13} and~\ref{fig14}, we show our results in this case for high-energy neutrino sources that are located at $z=1$, or that are distributed according to the star formation rate, in each case adopting the normal mass hierarchy, a sum of neutrino masses equal to 0.12 eV, $m_\chi\ll m_{Z'}$, and $g_{Z'}=5\times10^{-4}$, as motivated by the measured value of $g_{\mu}-2$. For the relatively large values of $g_{\chi}$ adopted in these figures, the attenuation feature can be more pronounced, providing a significantly better fit to the IceCube data.

Finally, we note that it is possible that the dark sector fermion in this scenario could be the dark matter of our universe. In Ref.~\cite{Holst:2021lzm}, the thermal relic abundance of the $\chi$ population was calculated, identifying regions of parameter space in which the current density of this particle could match the measured density of dark matter. We note that in order to generate an acceptable thermal relic abundance of dark sector fermions while also accommodating the measured value of $g_{\mu}-2$ in this scenario, we must require $m_{Z'} \gsim 20-30 \, {\rm MeV}$~\cite{Holst:2021lzm}. For this range of gauge boson masses, we expect neutrino scattering to lead to an attenuation feature at energies above $\sim 1 \, {\rm PeV}$, instead of in the sub-PeV range reported by IceCube. So while this scenario is not currently supported by the IceCube data, this is an interesting signature to look for in future high-statistics measurements of the  high-energy neutrino flux.

\section{Summary and Conclusions}\label{conclusions}
 
 Motivated by the measured value of $g_{\mu}-2$, we have considered in this study models with a broken $U(1)_{L_{\mu} - L_{\tau}}$ gauge symmetry, giving rise to a light gauge boson that couples to muons, taus, and thir respective neutrinos. Such a gauge boson, with a mass in the range of $m_{Z'} \sim 10-200 \, {\rm MeV}$ and with a coupling on the order of $g_{Z'} \sim(3-8) \times 10^{-4}$, could explain the measured value of the muon's magnetic moment while remaining consistent with all laboratory and cosmological constraints. Such a particle could also induce interactions between high-energy neutrinos and the neutrinos that make up the cosmic neutrino background, leading to spectral features in the diffuse high-energy neutrino spectrum, as measured by the IceCube Collaboration.
 
The spectrum of high-energy neutrinos reported by IceCube features what appears to be a dip at energies between $E_{\nu} \sim 0.2-1 \, {\rm PeV}$. While this spectral feature could plausibly arise from the properties of the sources themselves, we have taken it to motivate models in which neutrinos in this energy range are significantly attenuated by the scattering induced by a new gauge boson. In order for a gauge boson to produce an attenuation feature in this energy range, it must have a mass on the order of $m_{Z'} \sim \mathcal{O}(10)$ MeV.  In this study, we have identified a range of scenarios in which the presence of such a particle (with couplings chosen to resolve the $g_{\mu}-2$ anomaly) can significantly improve the fit to the IceCube data. For neutrino sources which follow well-motivated redshift distributions (such as those measured for BL Lacs Objects or which trace the star formation rate), the best fits are found for $m_{Z'} \sim 5-8 \, {\rm MeV}$. This range of masses, however, is in tension with cosmological measurements of the energy density in radiation, parameterized in terms of $N_{\rm eff}$~\cite{Escudero:2019gzq}. If the sources of the high-energy neutrinos are preferentially located at high redshifts, however, larger values of $m_{Z'} \sim 10-15 \, {\rm MeV}$ could accommodate IceCube's spectral feature, without leading to any unacceptable contributions to $N_{\rm eff}$.
   
In the future, we expect this situation to be substantially clarified for a number of reasons. First, the experimental uncertainties associated with the measurement of $g_{\mu}-2$ will be reduced considerably in the years ahead, along with the theoretical uncertainties on the Standard Model prediction for this quantity. Second, a combination of cosmological and laboratory measurements should allow us to determine the neutrino mass hierarchy, as well as the sum of the neutrino masses. Third, measurements of the diffuse spectrum of high-energy neutrinos will be further refined by IceCube and other neutrino telescopes~\cite{IceCube:2019pna}. Finally, future CMB experiments~\cite{CMB-S4:2016ple}, as well as line intensity mapping efforts~\cite{Karkare:2022bai,MoradinezhadDizgah:2021upg}, should be able to measure $\Delta N_{\rm eff}$ to a precision of $\sim 0.02-0.03$. In the parameter space of interest to this study, a detectable contribution to $N_{\rm eff}$ would be expected~\cite{Escudero:2019gzq}.

\begin{acknowledgments}  
 DH is supported by the Fermi Research Alliance, LLC under Contract No.~DE-AC02-07CH11359 with the U.S. Department of Energy, Office of Science, Office of High Energy Physics. JI is supported by a CNRS International collaboration program. This work has been done thanks to the facilities offered by the Univ. Savoie Mont Blanc - CNRS/IN2P3 MUST computing center. 
JI would like to thank Yoann Genolini for help with the setup of the computational strategy on GPUs, within the Labex Enigmass R\&D Booster program.
\end{acknowledgments}

\bibliography{Zprime}

\end{document}